# Nonlocal Electrical Detection of Reciprocal Orbital Edelstein Effect

Weiguang Gao[1,11], Liyang Liao[1,11], Hironari Isshiki[1], Nico Budai[1], Junyeon Kim[2,3], Hyun-Woo Lee[4,5], Kyung-Jin Lee[6], Dongwook Go[7], Yuriy Mokrousov[7,8], Shinji Miwa[1,9,10], and Yoshichika Otani[1,2,10]

[1] Institute for Solid State Physics, The University of Tokyo, Kashiwa, Chiba 277–8581, Japan

[2] Center for Emergent Matter Science, RIKEN, Wako, Saitama 351-0198, Japan

[3] National Institute of Advanced Industrial Science and Technology (AIST), Research Center for Emerging Computing Technologies, Tsukuba, Ibaraki 305-8568, Japan

[4] Department of Physics, Pohang University of Science and Technology, Pohang 37673, Korea.

[5] Asia Pacific Center for Theoretical Physics, Pohang 37673, Korea.

[6] Department of Physics, Korea Advanced Institute of Science and Technology, Daejeon 34141, Korea

[7] Institute of Physics, Johannes Gutenberg University Mainz, Mainz 55099, Germany.

[8] Peter Grünberg Institut and Institute for Advanced Simulation, Forschungszentrum Jülich and JARA, Julich 52428, Germany.

[9] CREST, Japan Science and Technology Agency (JST), Kawaguchi, Saitama 332-0012, Japan

[10] Trans-scale Quantum Science Institute, The University of Tokyo, Bunkyo-ku, Tokyo 113-0033, Japan

[11] These authors contributed equally.

e-mail: yotani@issp.u-tokyo.ac.jp

## Abstract

Spin-Orbitronics leverages the spin and orbital degrees of freedom in solids for information processing[1,2]. The orbital Edelstein effect[3–5] and orbital Hall effect[1,6–9], where the charge current induces a nonequilibrium orbital angular momentum, offer a promising method to manipulate nanomagnets efficiently using light elements[1,4–8]. Despite extensive research, understanding the Onsager reciprocity of orbital transport—fundamentally rooted in the second law of thermodynamics and time-reversal symmetry—remains elusive. In this study, we experimentally demonstrate the Onsager reciprocity of orbital transport in an orbital Edelstein system[4,10–12] by utilizing nonlocal measurements[13–23]. This method enables the precise identification of the chemical potential generated by orbital accumulation, avoiding the limitations associated with local measurements. Remarkably, we observe that the direct and inverse orbital-charge conversion processes produce identical electric voltages, confirming Onsager reciprocity in orbital transport. Additionally, we find that the orbital decay length, approximately 100 nm at room temperature, is independent of Cu thickness and decreases with lowering temperature, revealing a distinct contrast to spin transport behavior[15]. Our findings provide valuable insights into both the reciprocity of the charge-orbital interconversion and the nonlocal correlation of orbital degree of freedom, laying the ground for orbitronics devices with long-range interconnections.

## Introduction

Research on the dynamics and transport of different electronic degrees of freedom has driven advancements in physics, material science, and device engineering. Spin angular momenta (SAM) and its conversion, transport, and interaction with magnetization have been extensively studied in spintronics over the past decades. The spintronics community has utilized charge-to-spin conversion mechanisms, such as the well-known spin Hall effect[13,24] and spin Edelstein effect[25,26], to generate nonequilibrium SAM. Recently, the emerging field of orbitronics, which focuses on orbital angular momenta (OAM)[1,2,7,9], offers a complementary direction to the traditional reliance on SAM. Similar to charge-to-spin conversion, charge-to-orbital conversion occurs through the orbital Hall effect (OHE) and orbital Edelstein effect (OEE). Nonequilibrium OAM can be electrically induced in various light element systems with negligible spin-orbit coupling (SOC), such as Cr[27–29], Ti[1,30], Mn[28,31], $Al_2O_3$/Ru interface[32], oxides/Cu interface[10,11], and naturally oxidized Cu[12,33,34], as revealed by optical[1], terahertz emission[6], torque[3,10,11,29,32,35], magnetoresistance[28,33,36] and magnon[37] measurements, demonstrating the high efficiency of orbital torque and the broad selection of orbital source materials. However, these studies relied on local measurements, where the regions for OAM generation, distribution, and conversion overlap, potentially involving processes unrelated to OAM. This overlap could complicate the precise quantification of the interconversion between nonequilibrium OAM and charge current.

Nonlocal transport measurements[13–23] offer a promising approach to investigate various quantum transport phenomena, from diffusive spin transport in metal[13–20], unconventional spin conversion in moiré superlattices[38,39], to the quantum spin Hall[21] and quantum anomalous Hall effects[22,23]. In nonlocal transport devices, charge transport and nonequilibrium OAM can be spatially separated, enabling direct and isolated observation of the OAM distribution. Additionally, the nonlocal transport geometry naturally facilitates the measurement of the direct (DOEE) and inverse (IOEE) orbital Edelstein effect within a single device, allowing verification of Onsager's reciprocal relation, a principle of fundamental importance in the study of quantum transport.

In this study, we present measurements of lateral OAM distribution and the reciprocity of the orbital Edelstein effect using nonlocal transport devices consisting of $Al_2O_3$, Cu, and ferromagnets (FMs). The nonequilibrium OAM can reach approximately 100 nm along a lateral direction. This decay length remains unaffected by the Cu thickness, suggesting that OAM distribution in the lateral direction differs from that in the vertical direction, aligning with the formation of interfacial orbital Rashba states between Cu and oxides. The lateral decay length decreases with lower temperature, distinct from spin diffusion behavior. Crucially, all experimental results comply with Onsager's reciprocal relations. Our work offers significant insights into the nonlocal correlation behaviors of nonequilibrium OAM and expands the design possibilities for orbitronics devices.

**Measurement of orbital accumulation in nonlocal transport device structure**

We investigated the OAM distribution using a nonlocal lateral transport structure[13–20] (Fig. 1a), where the nonlocal resistance signals arise from DOEE and IOEE. In our samples, an oxidized Cu layer (denoted by $CuO_x$ hereafter) which is expected to be about 3 nm thick (Supplementary Section 1), is sandwiched by the $Al_2O_3$ and Cu due to the natural oxidization (method). In the direct measurement configuration (Fig. 1a), a charge current $I_c$ is applied to the $Al_2O_3/CuO_x/Cu$ nanowire which is oriented along the $y$-axis in Fig. 1a (denoted by $Cu_y$ nanowire hereafter). This current induces nonequilibrium OAM (**L**) polarized in the $x$-direction according to the OEE Hamiltonian, $H_{OEE} = \alpha_{OEE}(\mathbf{L} \times \mathbf{k}) \cdot \hat{\mathbf{z}}$, where $\alpha_{OEE}$ is the orbital Rashba coefficient, $\mathbf{k}$ is the wavevector, and $\hat{\mathbf{z}}$ is the unit vector along the $z$-axis. If a nonlocal orbital response occurs at the Cu/FM junction, it can be converted into spins through the SOC of the FM, generating an output voltage $V$ across the Cu/FM junction. We chose FM as the orbital converter instead of heavy metal, so that by varying the magnetization direction $\Phi$, we can separate the nonequilibrium OAM-induced signal from the bypass current-induced offset. The nonlocal signal as a function of the separation distance $d$ reflects the lateral distribution of the nonequilibrium OAM.

In the inverse measurement configuration, the processes are reversed according to Onsager's reciprocal relations (Fig. 1b). A charge current $I_c$ is injected into the $Al_2O_3/CuO_x/Cu$ nanowire which is oriented along the $x$-axis in Fig. 1b (denoted by $Cu_x$ nanowire hereafter) from the FM magnetized along the $x$-axis. This generates a nonequilibrium $x$-polarized OAM in the $Cu_x$ nanowire via the SOC of the FM, which is converted to a charge current through IOEE, resulting in the voltage signal $V_{IOEE}$. Spin current is also generated in this process, but the negligible SOC in Cu[16] should only lead to a tiny contribution to the voltage signal, as will be proved below.

We first demonstrate both direct and inverse measurements to validate the feasibility of measuring nonequilibrium OAM using nonlocal configuration. Figures 1c and 1d illustrate the typical direct ($R_{DOEE}$) and inverse ($R_{IOEE}$) orbital Edelstein resistance as a function of the magnetic field along the x-axis, observed in sample A (separation distance $d$ = 140 nm, Cu thickness $t_{Cu}$ = 40 nm and FM = Co$_{25}$Fe$_{75}$). $R_{DOEE}$ and $R_{IOEE}$ are defined as $R \equiv V/I_c$ where $R$ refers to both $R_{DOEE}$ and $R_{IOEE}$, $V$ is the measured voltage signal and $I_c$ is the applied current. Both $R_{DOEE}$ and $R_{IOEE}$ exhibit a linear response to the external magnetic field $\mathbf{B}_{ext}$ within the range of –0.5T to +0.5T, saturating once the magnitude of $\mathbf{B}_{ext}$ exceeds the saturation field $\mathbf{B}_{sat}$ (given by the anisotropic magnetoresistance curve, see Supplementary Section 2) of magnetization. Taking the difference of $R_{DOEE}$ and $R_{IOEE}$ at +$\mathbf{B}_{ext}$ and –$\mathbf{B}_{ext}$, we obtained a $2\Delta R_{DOEE}$ = 0.22 mΩ and $2\Delta R_{IOEE}$ = –0.22 mΩ. The absolute value of the signal is two orders of magnitude larger than the artifact coming from the stray field-induced Hall effect in Cu estimated via the COMSOL simulation (Supplementary Section 3).

As a typical way to provide evidence for nonequilibrium OAM[10,30,32,40,41], we conducted FM dependence experiments using Co$_{25}$Fe$_{75}$, Co$_{50}$Fe$_{50}$, and Ni$_{81}$Fe$_{19}$, as shown in Fig. 1e and 1f, where $2\Delta R$ is the overall change in nonlocal resistance. $|2\Delta R|$(Co$_{25}$Fe$_{75}$) is larger than $|2\Delta R|$(Co$_{50}$Fe$_{50}$), and $|2\Delta R|$(Ni$_{81}$Fe$_{19}$) is at least one order of magnitude smaller than them. A different measurement focusing on the local responses also confirms the FM dependence (measurement configurations and results are detailed in Supplementary Section 4). Consistent with previous studies[10,30,32,40,41], the FM dependent results identify the orbital characteristic of these responses. We attribute the strong FM dependence to the spin-orbit correlation $\langle \mathbf{L} \cdot \mathbf{S} \rangle^{FM}$ within FMs[40–42]. $\langle \mathbf{L} \cdot \mathbf{S} \rangle^{FM}$ near the Fermi level in FMs varies significantly with slight changes in FM composition, leading to $\langle \mathbf{L} \cdot \mathbf{S} \rangle^{Co_{25}Fe_{75}} > \langle \mathbf{L} \cdot \mathbf{S} \rangle^{Co_{50}Fe_{50}} > \langle \mathbf{L} \cdot \mathbf{S} \rangle^{Ni_{81}Fe_{19}}$ [40]. As the stray fields of the Co$_{25}$Fe$_{75}$ and the Ni$_{81}$Fe$_{19}$ are in the same order, the large difference between $|2\Delta R|_{(Co25Fe75)}$ and $|2\Delta R|_{(Ni81Fe19)}$ also excludes the stray field-induced artifacts in the nonlocal resistance. The spin-induced nonlocal resistance also cannot strongly depend on the FM type (see Supplementary Section 5, where $Co_{25}Fe_{75}$ and $Ni_{81}Fe_{19}$ show comparable nonlocal spin valve signals), suggesting that the observed $\Delta R$ is dominated by OAM.

Next, we analyzed the nonlocal resistance as a function of the separation distance $d$. The nonlocal resistance decreases exponentially with increasing $d$ (Fig. 1f), indicating that the OAM accumulation decays as the detection point moves farther from the charge current source. A significant signal persists up to $d \sim 350$ nm, corresponding to an exponential decay length on the order of 100 nm. While a trivial current bypass effect can generate a nonlocal response, as previously reported[43,44], both COMSOL simulations and analytical modeling estimate a decay length of approximately 47 nm (Supplementary Section 6). This is substantially shorter than the experimentally observed value of about 100 nm, thereby excluding the current bypass effect as the sole origin of the observed nonlocal response. Accounting for both OAM decay and bypass effects, we applied the following equation to evaluate the lateral decay length of orbital accumulation distribution:

$$2\Delta R = \frac{A}{\frac{1}{\lambda_o} - \frac{\pi}{W_x}}\left[\exp\left(-\frac{\pi d}{W_x}\right) - \exp\left(-\frac{d}{\lambda_o}\right)\right], \tag{1}$$

where $A$ is the fitting parameter for local processes, including the vertical OAM decay, charge-to-orbital and orbital-to-spin conversion, $\lambda_o$ is the lateral decay length of orbital accumulation, $W_x$ is the width of the Cu nanowire. As shown in Fig. 1e and 1f, fitting the data for $Co_{25}Fe_{75}$ (circles) and $Co_{50}Fe_{50}$ (diamonds) to Eq. (1) yields $\lambda_o \sim 100$ nm, regardless of the FM selection, showing that the nonlocal correlation is determined by the processes in the $Al_2O_3/CuO_x/Cu$ nanowire and unaffected by the Cu/FM junctions.

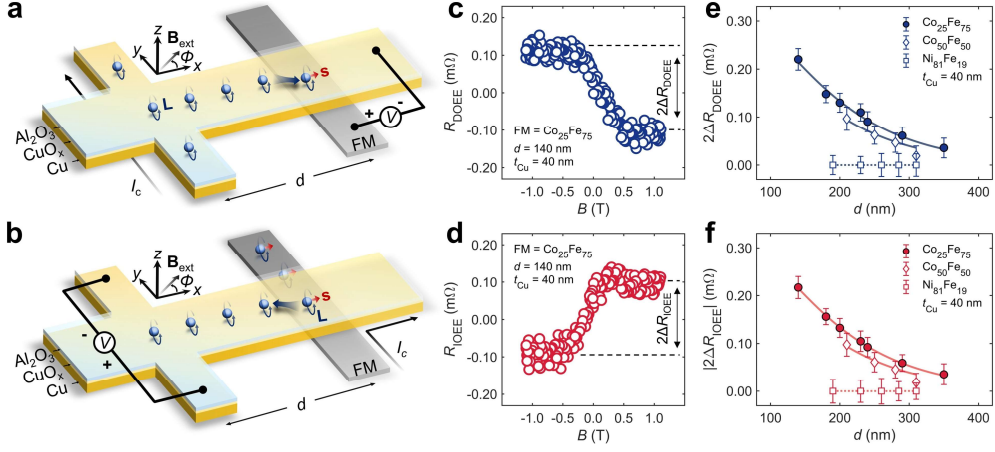

**Fig.1 | Schematic illustration of nonlocal transport measurements and verification of orbital response through ferromagnetic materials dependence experiments. a,** Nonlocal measurement configuration to observe DOEE (direct measurement). The nonequilibrium OAM are generated by charge current $I_c$ at the CuO$_x$/Cu interface through DOEE. The orbital accumulation then converts into SAM via SOC in FMs and induces a nonlocal response $V$. **b,** Nonlocal measurement configuration to observe IOEE (inverse measurement). A charge current brings nonequilibrium OAM from the FMs. The orbital accumulation then converts to charge current at the CuO$_x$/Cu interface through IOEE. In nonlocal measurement (**a** and **b**), the orbital generator and detector are sufficiently isolated in space with a separation distance $d$, allowing the measurement of nonlocal orbital response. **c, d,** Typical results of direct (**c**, $R_{DOEE}$) and inverse (**d**, $R_{IOEE}$) nonlocal orbital Edelstein resistance $R$ which is defined as $R \equiv V/I_c$. The results are observed in sample A with separation distance $d = 140$ nm, Cu thickness $t_{Cu} = 40$ nm and FM = Co$_{25}$Fe$_{75}$ (Methods) while the external magnetic field B$_{ext}$ swept along the hard axis of FM ($\Phi = 0°$) from $-1.25$ T to $1.25$ T. The signals are globally offset to position their center at $R = 0$ Ω. The double-headed arrows indicate the definition of $2\Delta R_{DOEE}$ and $2\Delta R_{IOEE}$, where $2\Delta R_{DOEE} = -2\Delta R_{IOEE} \approx 0.22$ mΩ. **e, f,** FM dependence results of $2\Delta R_{DOEE}$ (**e**) and $2\Delta R_{IOEE}$ (**f**). The solid curves represent the fitting of the data to Eq. (1), implying a long-range decay length of orbital accumulation $\lambda_o$ of about 100 nm regardless of the selection of

FMs. The dotted curves are guiding lines showing the value $R = 0\ \Omega$. The error bars indicate the noise level of $R$. All results are in solid agreement with Onsager's reciprocal relations.

Measuring nonequilibrium OAM using the nonlocal lateral transport structure demonstrates the feasibility of detecting OAM through the chemical potential, allowing the measurement of pure orbital accumulation. Onsager's reciprocal relations are then validated in the experiments: the relationship $2\Delta R_{\text{DOEE}} = -2\Delta R_{\text{IOEE}}$ holds across all measurements. This underscores that the OAM degree of freedom is an active and essential factor in electronic transport, a consideration overlooked in previous research.

**Angular dependence**

We then conducted an angular dependence experiment on sample A to clarify the direction of the OAM. As the angle $\Phi$ of $\mathbf{B}_{\text{ext}}$ to the $x$-axis was varied in-plane from 0° to 180°, $R_{\text{DOEE}}$ (Fig. 2a) and $R_{\text{IOEE}}$ (Fig. 2b) evolved accordingly. Notably, $2\Delta R_{\text{DOEE}}$ (Fig. 2c) and $2\Delta R_{\text{IOEE}}$ (Fig. 2d) show a strong correlation with the corresponding cosine curves $f(\Phi) = f_1 \cos \Phi$, where $f_1$ represents the value of $2\Delta R_{\text{DOEE}}$ or $2\Delta R_{\text{IOEE}}$ at 90°. Analogous to the angle dependence in spin measurements[14], this strongly suggests that we are detecting OAM projected along the magnetization direction. This trend further implies that the measured OAM is polarized along the $x$-direction, consistent with OEE, perpendicular to the applied charge current[4,45]. Minor deviations from the cosine curve, where the data points vertically shift towards $2\Delta R = 0$ $\Omega$, can be attributed to the magnetic anisotropy of FM nanowire (Supplementary Section 7).

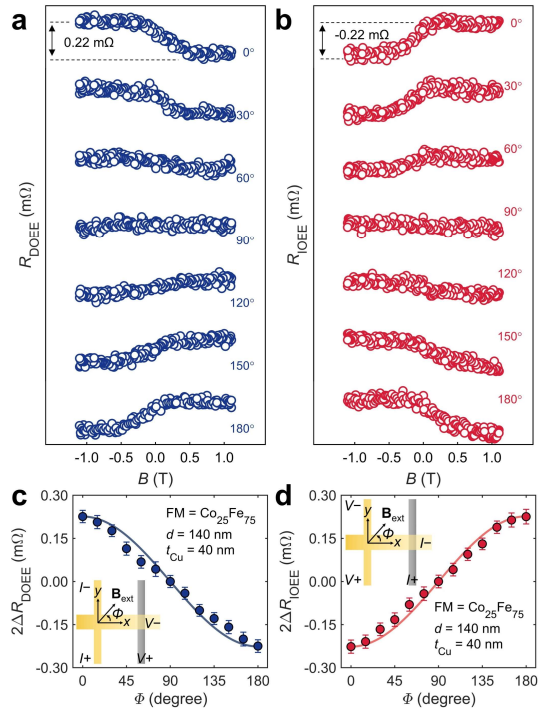

**Fig.2 | Angle dependence experiments. a, b** Nonlocal orbital Edelstein resistance $R_{\text{DOEE}}$ (**a**) and $R_{\text{IOEE}}$ (**b**) for sample A observed with magnetic field at various angle $\Phi$. **c, d,** $2\Delta R_{\text{DOEE}}$

and $2\Delta R_{\text{IOEE}}$ as a function of $\Phi$. The insets show the measurement configuration and the definition of angle $\Phi$. The results are measured under room temperature. The solid curves in (**c**) and (**d**) are guiding lines employing a cosine function $f(\Phi) = f_1 \cos\Phi$. The angle-dependent $2\Delta R_{\text{DOEE}}$ and $2\Delta R_{\text{IOEE}}$ well match the cosine function, indicating that $x$-polarized OAM are measured. The error bars in (**c**) and (**d**) represent the noise level of $R$. All results ascertain Onsager's reciprocal relations.

**Cu thickness dependence**

To reveal both the lateral and vertical distribution of the OAM in the $Al_2O_3/CuO_x/Cu$ nanowire system, we studied the Cu thickness dependence of the nonlocal signals (Fig. 3a), as summarized in Fig. 3b and 3c. Fitting $2\Delta R_{DOEE}$ (Fig. 3b) and $|2\Delta R_{IOEE}|$ (Fig. 3c) to Eq. (1), we observed that increasing $t_{Cu}$ causes only a slight change in $\lambda_o$ (Fig. 3d) but a significant reduction in $A$ (Fig. 3e). As the thickness of the oxidized Cu regions remains the same in all samples, due to the identical atmospheric exposure times, almost no dependence of $\lambda_o$ on the Cu thickness suggests that oxidized Cu supports the lateral nonlocal orbital response. Conversely, the thickness of unoxidized Cu depends on $t_{Cu}$. We thus assumed $A$ contains the information on the vertical orbital distribution as $A = A'_{Cu} \exp(-t_{Cu}/\lambda_o^z)$ where $A'_{Cu}$ is a thickness-independent fitting parameter, and $\lambda_o^z$ is the vertical decay length of OAM. The fitting (Fig. 3e) yields $\lambda_o^z \approx 25$ nm. The Cu thickness dependence in the local configuration shows consistent results (Supplementary Section 8).

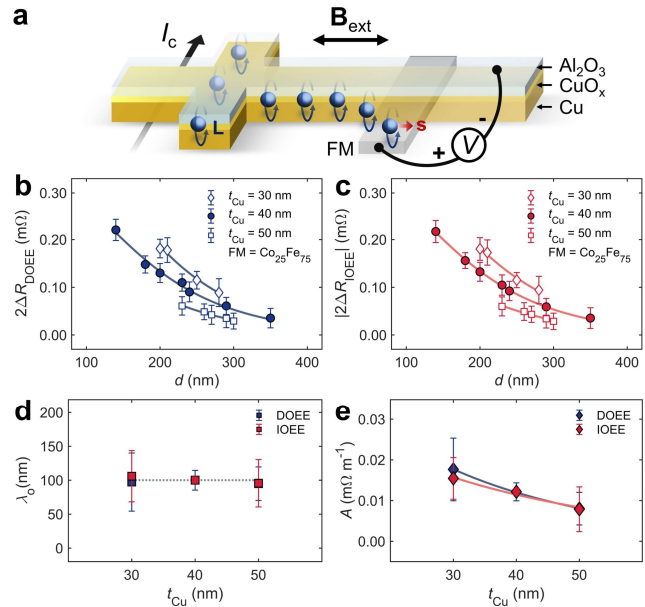

**Fig.3 | Cu thickness dependence experiments. a,** Schematic illustration of lateral and vertical OAM distribution and the role of oxidized and unoxidized Cu. The top Cu layer is

oxidized and homogenous, assisting the long-range orbital response; the bottom Cu layer remains unoxidized, exhibiting a short vertical decay length of orbital accumulation. **b, c,** Cu thickness dependence results of $2\Delta R_{DOEE}$ (**b**) and $2\Delta R_{IOEE}$ (**c**). The results are measured under room temperature. The solid curves represent the fitting of the data to Eq. (1). The error bars in (**b**) and (**c**) indicate the noise levels of $R$. **d,** $\lambda_o$ at various Cu thicknesses obtained from fitting where the values are almost constant. The dotted line marks the value of $\lambda_o = 100$ nm. **e,** Fitting result of parameter $A$. The solid curves show the exponential fitting for $A$. The vertical decay length of orbital accumulation ($\lambda_o^z \sim 25$ nm) is much smaller than the lateral one ($\lambda_o \sim 100$ nm), suggesting a distinction between lateral and vertical distribution of OAM. Error bars in (**d**) and (**e**) represent 95% confidence intervals from fitting results. All results agree with Onsager's reciprocal relations.

The distinction between $\lambda_o$ and $\lambda_o^z$ suggests different physical mechanisms governing the lateral and vertical OAM distribution in the oxide/Cu systems[10,11]. Previous studies on the vertical distribution of OAM[10,11,33] have highlighted the crucial role of oxidation. In oxidized Cu, the $3d$ electron shell is not fully occupied, orbital states allow electrons to host OAM. Such OAM can barely penetrate the unoxidized Cu region where the $3d$ electron shell is nearly fully occupied[46], leading to a short decay length. Meanwhile, the continuously oxidized Cu region with a certain thickness $t_{CuO}$ provides a uniform electronic band structure across the $x$-$y$ plane, supporting orbital accumulation associated with the large $\lambda_o$.

**Temperature dependence**

We next performed temperature-dependent experiments to gain deeper insight into the nonlocal orbital response. Figure 4a shows $R_{\text{DOEE}}$ at 300 K (black squares) and 50 K (green circles) in sample A ($d$ = 140 nm and FM = $Co_{25}Fe_{75}$). $R_{\text{DOEE}}$ signal is sizable at 300 K, but vanishing at 50 K, demonstrating that the nonlocal orbital response is diminishing at low temperature. To disentangle the local and nonlocal contributions during the temperature evolution, we also performed temperature dependent measurements in the local configuration, where the orbital response is measured by FM nanowires right below the orbital generation part, as shown in Fig. 4b. The local $R_{\text{DOEE}}^{(0)}$ at 300 K (black squares) is also larger than that at 50 K (green circles), although half of the signal survives at 50 K. These behaviors are in sharp contrast with the spin transport, characterized by $R_{\text{NLSV}}$ (Fig. 4c) in a nonlocal spin valve configuration, where the signal is larger at 50 K (green circles) than that at 300 K (black squares).

Figure 4d summarizes the temperature dependence of the local $2\Delta R_{\text{DOEE}}^{(0)}$ (green circles) and the nonlocal $2\Delta R_{\text{DOEE}}$ at different distances $d$ (other different colors). The corresponding raw data is presented in the Supplementary Section 9 (Fig. S12 and Fig. S13). The signal decreases at low temperatures in both local and nonlocal measurements. At 50 K, while the nonlocal signals drop to nearly zero, the local signals still retain a finite value, indicating that the local signal decreases more slowly than the nonlocal signal. Same temperature dependence is observed in the inverse measurement (Fig. S14), and no signal is observed at any temperature in the nonlocal orbital device with FM = $Ni_{81}Fe_{19}$ (Fig. S15). Meanwhile, the nonlocal spin signal $2\Delta R_{\text{NLSV}}$ increases with decreasing temperature (Fig.4e) and presents in both devices with FM = $Co_{25}Fe_{75}$ and $Ni_{81}Fe_{19}$, clearly indicating the different physics involved in the observed nonlocal orbital and the spin signals.

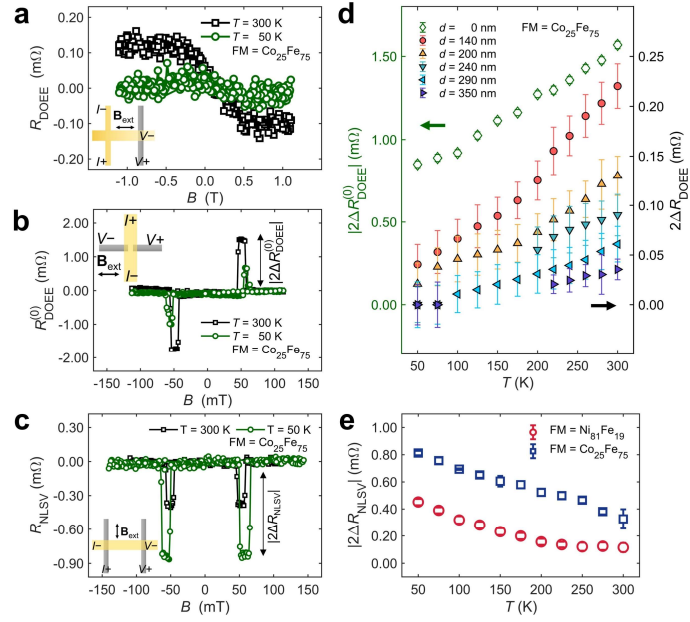

**Fig.4 | Temperature dependence experiments. a, b, c,** Typical signals for $R_{\text{DOEE}}$ (**a**) $R_{\text{DOEE}}^{(0)}$ (**b**) and $R_{\text{NLSV}}$ (**c**) at 300 K (black squares) and 50 K (green circles). $R_{\text{DOEE}}$ and $R_{\text{DOEE}}^{(0)}$ signals decrease with lower temperatures, while $R_{\text{NLSV}}$ signals increase as the temperature decreases. **d,** Temperature dependence results of $2\Delta R_{\text{DOEE}}^{(0)}$ (open diamond markers with green color, left $y$-axis) and $2\Delta R_{\text{DOEE}}$ (circle and triangle markers filled with other colors, right $y$-axis). The error bars indicate the noise levels of $R$. The results are measured in samples with FM = Co$_{25}$Fe$_{75}$. **e,** Summarized temperature dependence results of $2\Delta R_{\text{NLSV}}$ with Co$_{25}$Fe$_{75}$ (open blue squares) and Ni$_{81}$Fe$_{19}$ (open red circles). The error bars indicate the noise levels of $R$. The opposite temperature dependence on the behaviors of DOEE (**d**) and NLSV (**e**) implies distinct physics between OAM and SAM.

To isolate the contribution of the nonlocal process in the temperature dependence of $2\Delta R_{\text{DOEE}}$, we assume that the local process parameter $A$ for the nonlocal signal $2\Delta R_{\text{DOEE}}$ in Eq. (1) has the same temperature dependence as the local signal $2\Delta R_{\text{DOEE}}^{(0)}$. This leads to the relationship $A(T)/A(300\text{ K}) = \Delta R_{\text{DOEE}}^{(0)}(T)/\Delta R_{\text{DOEE}}^{(0)}(300\text{ K})$, allowing us to calculate the $A(T)$ values at different temperatures using $A(300$ K$)$ (Figs. 1e and 1f) and the temperature dependence of the local signal. We then fitted the distance $d$ dependence of $2\Delta R_{DOEE}$ (Fig. 5a) and $|2\Delta R_{IOEE}|$ (Fig. S16) at each temperature using Eq. (1) and $A(T)$ given above, so that $\lambda_o$ becomes the only fitting parameter, as summarized in Fig. 5b. Interestingly, $\lambda_o$ decreases as temperature drops, which contrasts with the typical behavior of spin transport in metals, where diffusion length is longer at lower temperatures[15]. The temperature dependent $\lambda_o$ also contrasts with the bypass effect which has no change at different temperature[43,44] (Supplementary Section 10). This suggests that different physical mechanisms mediate the nonlocality of OAM. Calculations have shown that orbital diffusion length in centrosymmetric systems is often below 10 nm[47]. Therefore, the nonlocality may not be linked to a diffusive orbital current, but rather due to the eigenstates in the Rashba band which host OAM by themselves: the Rashba effect locks the nonequilibrium OAM locally to the diffusive linear momentum[48,49] of the electrons. We propose a possible picture for the temperature dependence as shown in Fig. 5c. At low temperatures, the limited hopping between the localized states in the oxidized Cu cannot support orbital Rashba bands. The itinerant electrons have wavefunctions constrained in the unoxidized Cu layer without carrying OAM. The OAM propagation is more likely to occur in the high temperature region, where free carriers can be excited due to the thermal broadening of the localized states, and a conductive channel is formed at the oxidized Cu layer (Supplementary Section 11). As a result, the itinerant electrons have wavefunctions penetrating the oxidized Cu layer, creating hybridized states with OAM. Thus, continuous orbital Rashba bands are formed on the whole $Al_2O_3/CuO_x/Cu$ nanowire, and long-range orbital propagation is allowed.

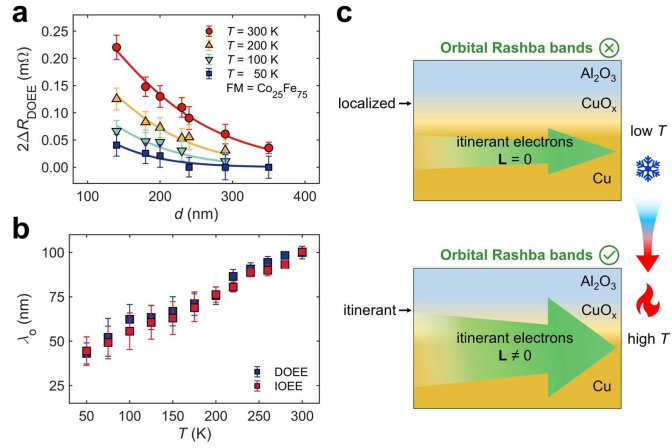

**Fig. 5 | Analysis on temperature-dependent orbital transport. a,** Fitting of the $2\Delta R_{\text{DOEE}}(T)$ data with Eq. (1). The solid curves represent the fitting results. The error bars indicate the noise levels of $R_{\text{DOEE}}$. **b,** Fitting result of $\lambda_o$ for $2\Delta R_{\text{DOEE}}$ (blue) and $2\Delta R_{\text{IOEE}}$ (red) as functions of temperature. The error bars indicate the 95% confidence intervals from fitting results. All the results obey Onsager's reciprocal relations. **c,** Schematic of the proposed mechanism for the temperature-dependent orbital propagation. At low temperatures (upper panel), hopping between the localized states in the oxidized Cu ($CuO_x$ layer) is limited, so the itinerant electrons are confined to the unoxidized Cu layer without carrying OAM ($\mathbf{L} = 0$). Thus, continuous orbital Rashba bands are not forming. At high temperature (lower panel), the localized states are thermally broadened, establishing the conductive channel in the $CuO_x$ layer. The electron wavefunctions are extended into the $CuO_x$ layer, creating hybridized states with OAM ($\mathbf{L} \neq 0$), which form the continuous orbital Rashba bands and allow the long-range orbital propagation.

**Conclusions**

Our study investigates the lateral distribution of nonequilibrium OAM and the reversible OEE by utilizing nonlocal response in an orbital Edelstein system composed of $Al_2O_3$, Cu, and ferromagnetic materials. The nonlocal response appears approximately 100 nm away from the injected electric current at room temperature, with its orbital nature confirmed by FM dependence and the polarization direction verified through angular dependence. The decay length $\lambda_o$ of orbital accumulation is unaffected by Cu thickness, suggesting that the oxidized Cu region assists the long-range nonlocal orbital response. In contrast to the spin diffusion length, $\lambda_o$ reduces with temperature decreasing, indicating that mechanisms other than diffusive orbital current mediate OAM transport. Importantly, in all experiments, the orbital and charge accumulation induced by DOEE and IOEE are observed in the same devices and are mathematically equivalent, consistent with Onsager's reciprocal relations. Our work provides clear insights into the intrinsic reciprocal relationship of orbital effects and reveals a long-range lateral correlation of OAM accumulation, paving the way for interconnectable orbitronics devices capable of operating over long distances.

**Acknowledgement**

We thank Thierry Valet for fruitful discussions and helpful comments. We also thank Shoya Sakamoto and Jieyi Chen for assistance. L.L. would like to thank the support from JSPS through "Research program for Young Scientists" (no. 23KJ0778). J.K. and Y.O. appreciate the financial support from Japan Society for the Promotion of Science (JSPS) KAKENHI (Grants No. JP19K05258, No.JP23K04574, and No. JP19H05629). H.-W.L. was supported by the Samsung Science and Technology Foundation (BA-1501-51) and the National Research Foundation of Korea (NRF) grant funded by the Korean government (MSIT) (No. RS-2024-00410027). K.-J.L. was supported by the National Research Foundation of Korea (NRF) grant funded by the Korean government (MSIT) (No. 2022M3H4A1A04098811). D.G. and Y.M. gratefully acknowledge financial support by the Deutsche Forschungsgemeinschaft (DFG, German Research Foundation) - TRR 288/2 - 422213477 (project B06), TRR 173/2 - 268565370 (project A11), and funding from the European Union's HORIZON EUROPE, under the grant agreement No 101129641.

**Competing interests**

The authors declare no competing interests.

**Author contribution**

Y.O., H.I., W.G. and J.K. designed the experiment. W.G. performed the experiment with the help of H.I. and S.M.. L.L., W.G., J.K. and Y.O. analyzed the data. N.B. and W.G. performed COMSOL simulations. L.L. developed the theoretical models with the help of H.-W.L.. D.G., Y.M. and K.-J.L. asserted the models. Y.O. supervised the project. L.L., W.G. and Y.O. wrote the paper with input from all the authors. All the authors reviewed and revised the paper. W.G. and L.L. contributed equally to the whole work.

**Additional information**

**Supplementary information** Supplementary information is available in the following part of this file.

**Correspondence and requests for materials** should be addressed to Yoshichika Otani.

**Methods**

**Sample fabrication**

Our devices were microfabricated on $SiO_2/Si$ substrates through electron beam lithography on polymethyl-methacrylate (PMMA) resist, develop, deposition, and lift-off processes. The ferromagnet (FM) nanowires were electron-beam evaporated on $SiO_2/Si$ substrates. Before the Cu deposition, the FM surface was Ar-ion milled carefully to obtain a transparent interface between FM and Cu nanowires. The cross-shaped Cu nanowires were deposited using a Joule heating evaporator. The samples were exposed to the atmosphere for 10 minutes before $Al_2O_3$ deposition. The $Al_2O_3$ capping layers were only deposited on the Cu nanowires by electron beam deposition. All the fabrication processes were performed at room temperature, and the base pressure of deposition was lower than $5.0 \times 10^{-8}$ Torr, and the pressure during deposition was lower than $1.0 \times 10^{-7}$ Torr.

All samples were fabricated in an identical and carefully designed geometry with specific variations explicitly noted. FM nanowires utilized $Co_{25}Fe_{75}$, with dimensions of 20 nm in thickness ($t_{FM}$) and 100 nm in width ($w_{FM}$). For the FM material dependence experiment, alternatives, $Co_{50}Fe_{50}$ or $Ni_{81}Fe_{19}$, were used for devices with the same geometry described above. The Cu nanowires lying on the $y$-axis are 100 nm wide, and the one lying on the $x$-axis is 150 nm wide. The thickness of Cu ($t_{Cu}$) is 40 nm, while in the Cu thickness dependence experiment, we also set $t_{Cu} = 30$ nm and 50 nm. The thickness of the $Al_2O_3$ capping layer ($t_{Al2O3}$) is 15 nm for all samples.

**Electrical Measurements**

All the electric transport measurements were carried out in a He flow cryostat using the lock-in measurement technique. Electrical contacts to devices were made with wire bonding with Al wires. The applied alternating current was carefully calibrated to 500 μA for the samples with different Cu thicknesses. The in-plane sweeping external magnetic field $\mathbf{B}_{ext}$ is applied to the hard axis of the FM wire for nonlocal measurement. The $\mathbf{B}_{\text{ext}}$ was rotated from the $x$-axis by the angle $\Phi$ ($0° \sim 180°$) for the angle dependence experiment.

# Supplementary Information

## Nonlocal Electrical Detection of Reciprocal Orbital Edelstein Effect

Weiguang Gao[1,11], Liyang Liao[1,11], Hironari Isshiki[1], Nico Budai[1], Junyeon Kim[2,3], Hyun-Woo Lee[4,5], Kyung-Jin Lee[6], Dongwook Go[7], Yuriy Mokrousov[7,8], Shinji Miwa[1,9,10], and Yoshichika Otani[1,2,10]

[1] Institute for Solid State Physics, The University of Tokyo, Kashiwa, Chiba 277–8581, Japan

[2] Center for Emergent Matter Science, RIKEN, Wako, Saitama 351-0198, Japan

[3] National Institute of Advanced Industrial Science and Technology (AIST), Research Center for Emerging Computing Technologies, Tsukuba, Ibaraki 305-8568, Japan

[4] Department of Physics, Pohang University of Science and Technology, Pohang 37673, Korea.

[5] Asia Pacific Center for Theoretical Physics, Pohang 37673, Korea.

[6] Department of Physics, Korea Advanced Institute of Science and Technology, Daejeon 34141, Korea

[7] Institute of Physics, Johannes Gutenberg University Mainz, Mainz 55099, Germany.

[8] Peter Grünberg Institut and Institute for Advanced Simulation, Forschungszentrum Jülich and JARA, Julich 52428, Germany.

[9] CREST, Japan Science and Technology Agency (JST), Kawaguchi, Saitama 332-0012, Japan

[10] Trans-scale Quantum Science Institute, The University of Tokyo, Bunkyo-ku, Tokyo 113-0033, Japan

[11] These authors contributed equally.

e-mail: yotani@issp.u-tokyo.ac.jp

# Table of contents

**Section 1.    The distribution of oxygen element in Al$_2$O$_3$/CuO$_x$/Cu nanowire nanowire**

Due to the fabrication process, the Cu layer in our sample was exposed to air for 10 minutes before Al$_2$O$_3$ deposition, likely resulting in a thin oxidized Cu (CuO$_x$) layer between Al$_2$O$_3$ and Cu. Scanning transmission electron microscopy (STEM) and energy-dispersive X-ray spectrometry (EDX) were used to characterize oxygen distribution (Fig. S1a). Fig. S1b presents the EDX line near the Al$_2$O$_3$/Cu interfaces. These analyses reveal an approximately 3 nm thick CuO$_x$ layer at the Al$_2$O$_3$/Cu interface showing consistency with previous results[1–3].

**Fig. S1**

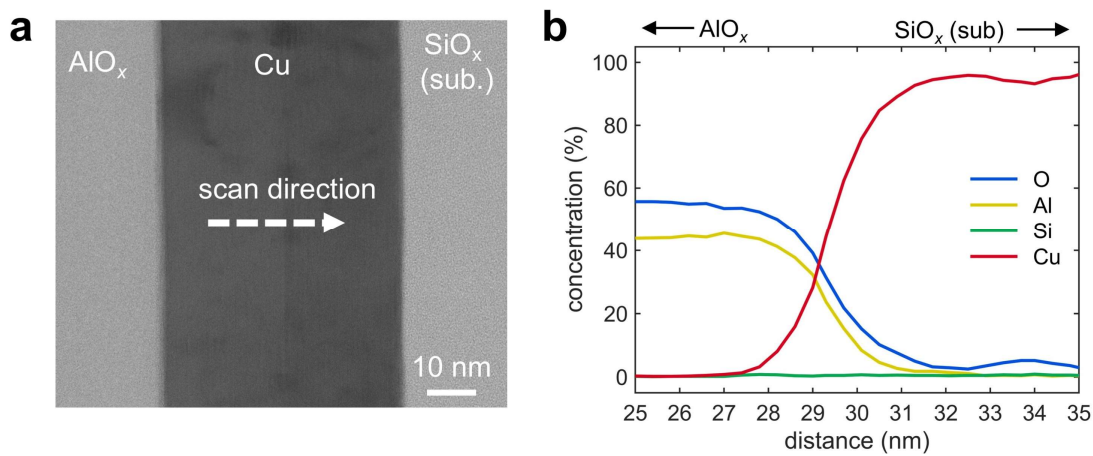

**Fig. S1 | STEM and EDX analysis results. a,** STEM image for Al$_2$O$_3$/CuO$_x$/Cu sample, representing the observed area for the EDX analysis. **b,** O, Al, Si, Cu atom EDX line profile near the Al$_2$O$_3$/Cu interface. The sample was deposited on SiO$_2$/Si substrate.

**Section 2.       The anisotropic magnetoresistance of ferromagnets**

The anisotropic magnetoresistances (AMR) were measured to estimate the saturation field of magnetization by using the four-probe measurements method and applying an external magnetic field along the hard axis of ferromagnets (FMs). The AMR signals of $Co_{25}Fe_{75}$, $Co_{50}Fe_{50}$, and $Ni_{81}Fe_{19}$ are shown in Fig. S2a, Fig. S2b, and Fig. S2c, respectively. In nonlocal transport measurements, the amplitude of external magnetic fields is set to 1.25T, which is large enough to saturate the magnetizations.

**Fig. S2**

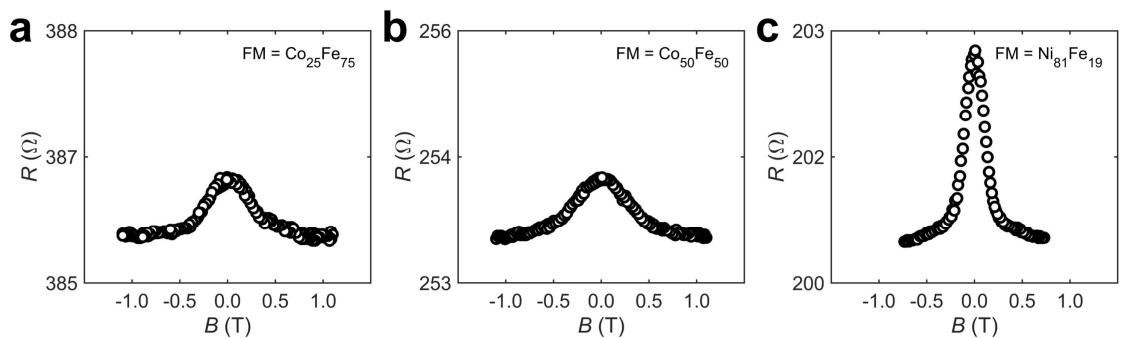

**Fig. S2 | Measurement of anisotropic magnetoresistance (AMR) of different FMs. a, b, c,** The typical AMR result of $Co_{25}Fe_{75}$ (**a**), $Co_{50}Fe_{50}$ (**b**) and $Ni_{81}Fe_{19}$ (**c**).

**Section 3.    COMSOL simulation for Hall voltage induced by stray field and bypass current**

**The bypass current in nonlocal measurement**

Consider our direct orbital Edelstein effect (DOEE) measurement (direct measurement) structure (see Fig. 1a in main text). When a charge current density $j_c$ flows in the longitudinal $Al_2O_3/CuO_x/Cu$ nanowire, which lies along the $y$-axis (denoted as $Cu_y$ nanowire hereafter), it shunts to the transverse $Al_2O_3/CuO_x/Cu$ nanowire which lies on $x$-axis (denoted as $Cu_x$ nanowire hereafter), as shown in Fig. S3a. The magnitude of bypass charge current density $j_{by}$ can be sizeable when the separation distance $d$ between generator and detector is comparable to the width of $Cu_x$ nanowire $W^{4,5}$ and induce a significant error in our evaluation of the lateral decay length of orbital accumulation $\lambda_o$.

The bypass current can induce both $y$-polarized and $x$-polarized orbital angular momenta (OAM), as shown in Fig. S3b. Although the $y$-polarized orbital accumulation can be generated by the $x$-component of bypass current density $j_{by}^x$ through DOEE, the positive (induced by $j_{by}^{+x}$) and negative polarized component (induced by $j_{by}^{-x}$) neutralize each other, leaving no net impact. Therefore, only the $y$-component of bypass current density $j_{by}^y$ give rise to the net contribution to the detected signals in our experiments, as shown in Fig. S3c. The detail discussion on bypass effect is in Section 6.

**Fig. S3**

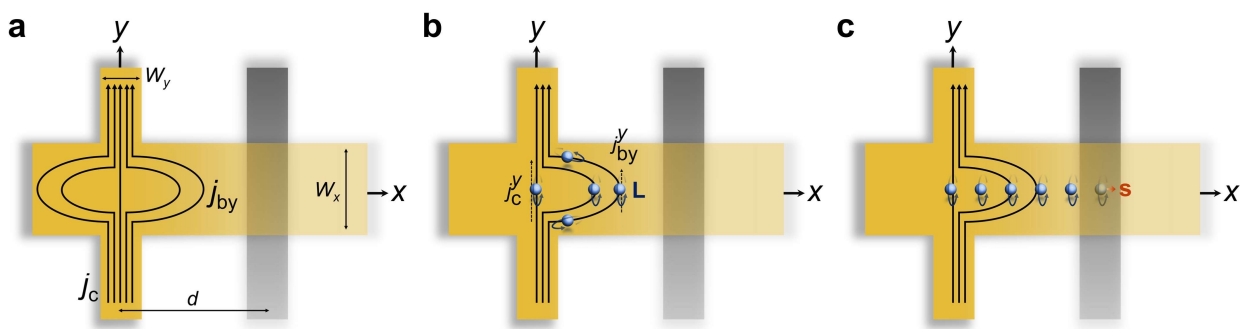

**Fig. S3 | The bypass effect of electric current in nonlocal measurement. a,** The systematic illustration for the combined contribution of bypass effect of charge current and the lateral distribution of OAM. The charge current applied to the $Cu_y$ nanowire shuts to the $Cu_x$ nanowire. **b,** Subsequently, the bypass current density induces nonequilibrium OAM at location $x$ (middle panel). **c,** Lastly, all orbital accumulation undergoes a decay before inducing a nonlocal orbital response in FM (right panel). Thus, the signal (nonlocal OEE resistance) is proportional to the volume integration of the final orbital accumulation density by considering the above three stages.

### The stray field in nonlocal measurements

Given the short channel length between $Cu_y$ and FM nanowires, in the neighbor where stray field $\mathbf{B}_{stray}$ and bypass current density $\mathbf{j}_{by}$ orthogonally coexist, a Hall electric field can be induced. We conducted an analytical discussion and COMSOL simulation to study the artifact induced by the Hall voltage. Given that the length of FM nanowires is much longer than their width and thickness, we assume that the stray field has no $y$-component and is homogenous along the $y$-direction in the analytical discussion. Since both $\mathbf{B}_{stray}$ and $\mathbf{j}_{by}$ are inhomogeneous, the Hall effect contribution is rather complicated. We hence focus on the leading order contribution from the Hall effect, whose existence relies on the homogeneous parts of the $\mathbf{B}_{stray}$ and $\mathbf{j}_{by}$ within a cuboid of the Cu nanowire and ignore the higher order contribution that relying on the inhomogeneity of the $\mathbf{B}_{stray}$ and $\mathbf{j}_{by}$ within a cuboid of the Cu nanowire.

### Analytical discussion of Hall effect induced by stray field and bypass current

Considering the direct measurement, the detected signal $V$ reflects the potential difference between FM terminal and the right arm of $Cu_x$ nanowire (Fig. S4a), which include the vertical potential difference ($V_z$) at the Cu/FM junction and $x$-direction potential ($V_x$) difference between the Cu/FM junction and right end of $Cu_y$ nanowire. Thus, only in the region B where $\mathbf{B}_{\text{stray}}$ and $\mathbf{j}_{\text{by}}$ satisfies $E_z \sim -j_{\text{by}}^y B_{\text{stray}}^x$ (Fig. S4b) and in the region A where $\mathbf{B}_{\text{stray}}$ and $\mathbf{j}_{\text{by}}$ satisfies $E_x \sim j_{\text{by}}^y B_{\text{stray}}^z$ (Fig. S4c), the Hall effect contributes to the detected signal $V$.

Considering the inverse measurement, the measured signal $V$ reflects the $y$-direction potential difference ($V_y$) between two ends of $\text{Cu}_y$ nanowires (Fig. S5a). Thus, only in the region B where $\mathbf{B}_{\text{stray}}$ and $\mathbf{j}_c$ satisfies $E_y \sim j_c^z B_{\text{stray}}^x$ (Fig. S5b) and in the region A where $\mathbf{B}_{\text{stray}}$ and $\mathbf{j}_c$ satisfies $E_y \sim -j_c^x B_{\text{stray}}^z$ (Fig. S5c), the Hall effect contributes to the detected signal $V$.

We briefly analyze the region C, D and E, where stray field and bypass current density orthogonally coexist but no contribution is expected. In the direct measurement, region C, D and E allow $E_x \sim j_c^y B_{\text{stray}}^z$, but the measurement circuit along the FM terminal and the right arm of $\text{Cu}_x$ nanowire cannot pick up this electric field. In the inverse measurement, region C, D and E have $E_y \sim j_c^x B_{\text{stray}}^z$, but given that $B_{\text{stray}}^z$ is homogeneous along the $y$-direction and the net current $I_x$ is zero, the net contribution of this electric field component to the measured signal is vanishing. Note that the reciprocity law requires a region to contribute to the signal in both of the measurements, or neither of the measurements, so that the A and B regions contribute to both the direct and inverse measurements and C, D, E have vanishing contribution (at the leading order) to both of them. We summarized the Hall effect produced by $\mathbf{B}_{\text{stray}}$ and $\mathbf{j}_{\text{by}}$ and their contribution to the measured signal in table 1 (direct measurement) and table 2 (inverse measurement) below.

**Table 1 | Hall effect in direct measurement.**

| Direct | $B_{\text{stray}}^x$ | $B_{\text{stray}}^y$ | $B_{\text{stray}}^z$ |
|---|---|---|---|
| $j_{\text{by}}^x$ | ●   Non-orthogonal. | ●   $B_{\text{stray}}^y = 0$ | ●   $\text{E}_y \sim -j_{\text{by}}^x B_{\text{stray}}^z$ <br> ●   Not detected by direct measurement. |
| $j_{\text{by}}^y$ | ●   $E_z \sim -j_{\text{by}}^y B_{\text{stray}}^x$ | ●   Non-orthogonal. <br>     $B_{\text{stray}}^y = 0$ | ✓   $E_x \sim j_{\text{by}}^y B_{\text{stray}}^z$ |
| $j_c^z$ | ●   $j_c^z$ does not exist. | ●   $j_c^z$ does not exist. <br> ●   $B_{\text{stray}}^y = 0$ | ●   Non-orthogonal. <br> ●   $j_c^z$ does not exist. |
| Region | Hall effect | | |
| A | $E_x \sim j_{\text{by}}^y B_{\text{stray}}^z$. | | |
| B | $E_z \sim -j_{\text{by}}^y B_{\text{stray}}^x$. | | |
| C | $E_x \sim j_c^y B_{\text{stray}}^x$. Yet the $E_x$ is not in the measurement circuit. | | |
| D | $E_x \sim j_c^y B_{\text{stray}}^z$. Yet the $E_x$ is not in the measurement circuit. | | |
| E | $E_x \sim j_c^y B_{\text{stray}}^z$. Yet the $E_x$ is not in the measurement circuit. | | |

**Table 2 | Hall effect in inverse measurement.**

| Inverse | $B_{\text{stray}}^x$ | $B_{\text{stray}}^y$ | $B_{\text{stray}}^z$ |
|---|---|---|---|
| $j_{\text{by}}^x$ | ●   Non-orthogonal. | ●   $B_{\text{stray}}^y = 0$ | ✓   $\text{E}_y \sim -j_{\text{by}}^x B_{\text{stray}}^z$ |
| $j_{\text{by}}^y$ | ●   $E_z \sim -j_{\text{by}}^y B_{\text{stray}}^x$ <br> ●   Not detected by inverse measurement. | ●   Non-orthogonal. <br>     $B_{\text{stray}}^y = 0$ | ●   $E_x \sim j_{\text{by}}^y B_{\text{stray}}^z$ <br> ●   Not detected by inverse measurement. |
| $j_c^z$ | ✓   $E_y \sim j_c^z B_{\text{stray}}^x$ | ●   $B_{\text{stray}}^y = 0$ | ●   Non-orthogonal. |
| Region | Hall effect | | |
| A | $E_y \sim -j_{\text{by}}^x B_{\text{stray}}^z$. | | |
| B | $E_y \sim j_c^z B_{\text{stray}}^x$. | | |
| C | $E_y \sim -j_{\text{by}}^x B_{\text{stray}}^z$. Yet no leading order contribution because $\int j_{\text{by}}^x dy dz = 0$. | | |
| D | $E_y \sim -j_{\text{by}}^x B_{\text{stray}}^z$. Yet no leading order contribution because $\int j_{\text{by}}^x dy dz = 0$. | | |
| E | $E_y \sim -j_{\text{by}}^x B_{\text{stray}}^z$. Yet no leading order contribution because $\int j_{\text{by}}^x dy dz = 0$. | | |

**COMSOL simulation**

Having established the symmetric analysis of the Hall effect in each part of the devices, we hereby study the Hall voltage induced by stray field and bypass current in both direct and inverse measurements through COMSOL simulations. We first estimated the Hall voltage in the direct measurement using the simulated stray field and bypass current density at two specific point A and B, corresponding to the A and B regions in the previous section. Point A is located 75 nm to the right of the FM nanowire, while point B is directly above the FM nanowire. Both points are positioned along the center axis of the $Cu_x$ nanowire.

At point A, the $z$ component of the stray field is 0.083 T, and the bypass current density is $2.3 \times 10^9$ A/m², resulting in a Hall electric field 0.01 A/m along the x direction (taking a Hall coefficient[6] of Cu $R_H = 5.3 \times 10^{-11} \mathrm{m}^3 \mathrm{A}^{-1} \mathrm{s}^{-1}$). Considering a characteristic distance of 100 nm in the nanowire device, the Hall voltage is 1 nV. At point B, the $x$-component of the stray field is $-0.14$ T and the bypass current density is $1.0 \times 10^{10}$ A/m², resulting in a Hall electric field 0.074 A/m along the $z$ direction. Using the thickness 40 nm of the Cu nanowire, the Hall voltage is 3 nV. Hence, the Hall effect contribution is at the order of 1nV, which is approximately two orders of magnitude smaller than the experimentally measured signal (100 nV).

To achieve a more precise estimation of the Hall voltage, we refined our approach by dividing the original nonlocal structure into $n$ hypothetical finite volume elements (here we chose $n = 4$). For each volume element, we calculated the average stray field and bypass current, which were then used to determine the Hall voltage within that specific element (Fig. S4d and Fig. S4e for direct measurement, and Fig. S5d and Fig. S5e for inverse measurement). By summing up the contributions from all the volume elements, we obtained a more accurate estimation of the overall Hall voltage. This method, as the COMSOL picture illustrated in Fig. S6a and Fig. S6b for direct and inverse measurement, provides a significant improvement in accuracy compared to a single-point estimation in COMSOL. Using this refined method, the estimated Hall voltage was found to be less than 3 nV under the applied current of 500 μA. Based on this value, the Hall effect-induced signal is calculated to be approximately 0.006 mΩ. In contrast, the experimentally measured signal is 0.22 mΩ, which is about two orders of magnitude larger than the calculated Hall effect contribution. This significant discrepancy strongly suggests that the Hall voltage induced by the combined effects of the stray field and bypass current has a negligible contribution to the measured signal in our experiments. Therefore, we conclude that the influence of the stray field can be effectively excluded from the observed signal.

**Fig. S4**

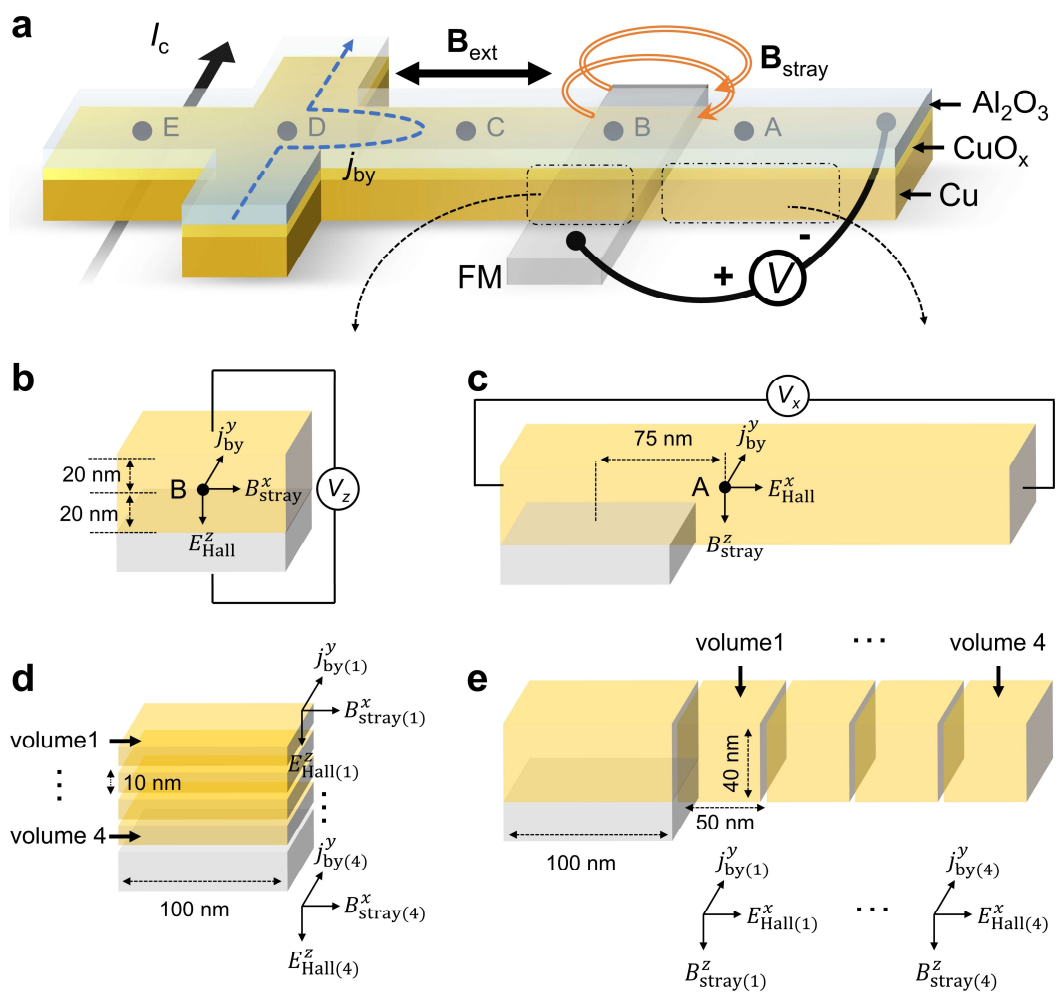

**Fig. S4 | Hall voltage induced by stray field and bypass current in direct measurement. a,** Schematic illustration of the stray field and bypass current in the direct measurement setup. The bypass current density ($j_{by}$) is represented by blue dashed curves, while the stray field ($\mathbf{B}_{stray}$) is shown in orange curves. Specific points along the centerline of the Cu$_x$ nanowire, labeled as A, B, C, D, and E, are highlighted for further analysis. **b,** $\mathbf{B}_{stray}$ and $j_{by}$ at the region near B point (above Cu/FM junction). A negative $z$-direction Hall electric field is induced, which can be detected by voltmeter. **c,** $\mathbf{B}_{stray}$ and $j_{by}$ at the region near A point (on the right of Cu/FM junction). An $x$-direction Hall electric field is induced, which can be detected by voltmeter. **d, e,** The analysis method for COMSOL simulation. The bulk of Cu above FM (**d**) and that to the right of FM (**e**) is divided into 4 volume elements each. For each volume element, $\mathbf{B}_{stray}$ and $j_{by}$ are simulated and averaged in each volume while the Hall voltage is calculated. This method provides a more accurate estimation by summing the contributions from all volume elements, compared to calculations based solely on a single point.

**Fig. S5**

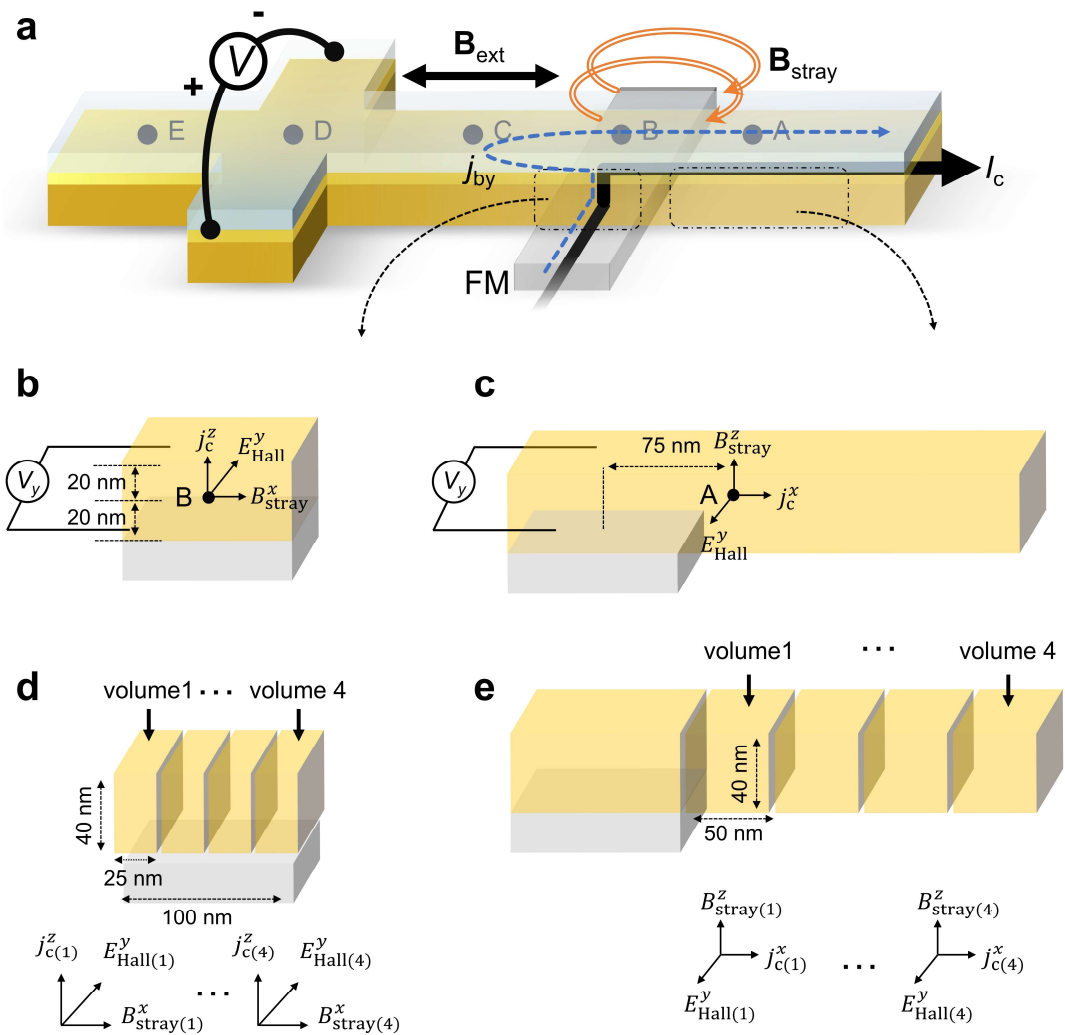

**Fig. S5 | Hall voltage induced by stray field and bypass current in inverse measurement. a,** Schematic illustration of the stray field and bypass current in the inverse measurement setup. The bypass current density ($j_{by}$) is represented by blue dashed curves, while the stray field ($\mathbf{B}_{stray}$) is shown in orange curves. Specific points along the centerline of the Cu$_x$ nanowire, labeled as A, B, C, D, and E, are highlighted for further analysis. **b,** $\mathbf{B}_{stray}$ and $j_{by}$ at the region near B point (above Cu/FM junction). A $y$-direction Hall electric field is induced, which can be detected by voltmeter. **c,** $\mathbf{B}_{stray}$ and $j_{by}$ at the region near A point (on the right of Cu/FM junction). A $y$-direction Hall electric field is induced, which can be detected by voltmeter. **d, e,** The analysis method for COMSOL simulation. The bulk of Cu above FM (**d**) and that to the right of FM (**e**) is divided into 4 volume elements

each. For each volume element, $\mathbf{B}_{stray}$ and $j_{by}$ are simulated and averaged in each volume while the Hall voltage is calculated. This method provides a more accurate estimation by summing the contributions from all volume elements, compared to calculations based solely on a single point.

**Fig. S6**

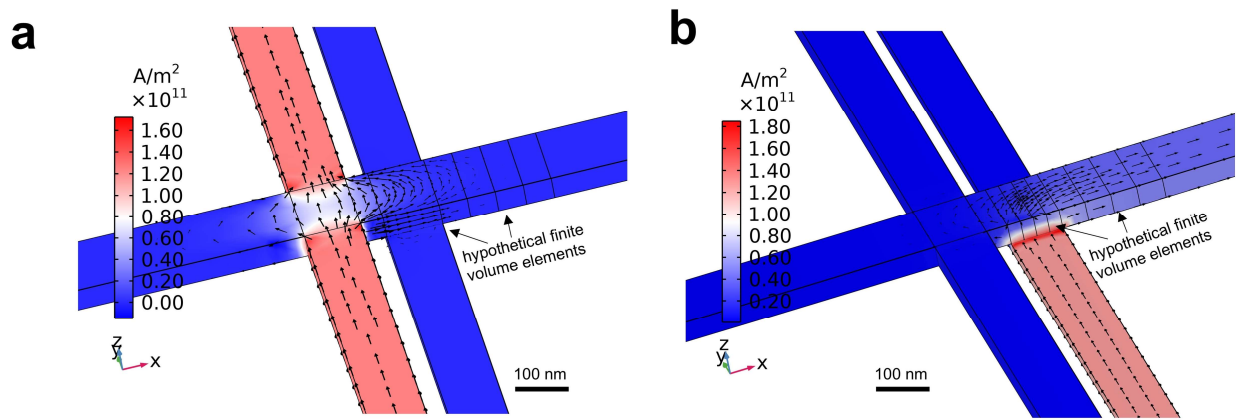

**Fig. S6 | The illustration of COMSOL simulation on Hall voltage. a, b,** COMSOL simulation of direct (**a**) and inverse (**b**) measurement setup. Hypothetical volume elements are depicted, while $\mathbf{B}_{stray}$ is omitted for clarity. For each volume element, $\mathbf{B}_{stray}$ and $j_{by}$ are simulated and averaged, followed by the calculation of the Hall voltage within each volume element. The Hall voltages from all elements are then summed to determine the total Hall voltage, which contributes negligibly to the measured signal.

**Section 4.    Orbital accumulation measured in local transport structure**

To explore solely the local distribution of orbital accumulation, we conducted the measurements exploiting a local transport structure[7] consisting of an $Al_2O_3/CuO_x/Cu$ nanowire (lays in the $x$-axis) on top of two separate FM nanowires (lays in the $y$-axis). The FM electrodes are patterned differently to have different switching fields. For the DOEE measurement (Fig. S7a), the nonequilibrium OAM are induced by charge current $I_c$ via DOEE (charge-to-orbital conversion). In FMs, the orbital accumulation converts to spin accumulation (orbital-to-spin conversion) due to the spin-orbit coupling (SOC) of FMs and shifts the spin chemical potential in FMs. The difference of chemical potentials in FM gives rise to the output voltage $V$. In contrast, in the inverse orbital Edelstein effect (IOEE) measurement (Fig. S7b), all processes are reversed. The nonequilibrium spin accumulation induced by charge current $I_c$ convert to orbital accumulation due to SOC in FM (spin-to-charge conversion). The orbital accumulation converts to the charge current via IOEE (orbital-to-charge conversion), inducing a charge current and causing an output voltage of $V$. However, the spin current remains unaffected because of the lack of SOC in the $Al_2O_3/CuO_x/Cu$ nanowire. In both measurement configurations, the sweeping external magnetic field $\mathbf{B}_{ext}$ is applied along the easy axis of FMs ($y$-axis). The local direct ($R^{(0)}_{DOEE}$) and inverse ($R^{(0)}_{IOEE}$) orbital Edelstein resistance are defined as $R^{(0)} \equiv V/I_c$, where $R^{(0)}$ refers to both $R^{(0)}_{DOEE}$ and $R^{(0)}_{IOEE}$. Absolute value $|2\Delta R^{(0)}_{DOEE}|$ and $|2\Delta R^{(0)}_{IOEE}|$ refer to the overall change of $R^{(0)}_{DOEE}$ and $R^{(0)}_{IOEE}$. In local transport measurement, the orbital generator and detector are vertically separated in space, which allows the detection of pure vertical distribution of OAM.

The samples for local transport measurement were microfabricated on $SiO_2/Si$ substrates through the electron beam lithography on polymethyl-methacrylate (PMMA) electron beam photoresist, develop, deposition, and lift-off processes. All devices share the same design with specific variations explicitly noted. The 100 nm wide and 20 nm thick FM nanowire pairs with 50 nm gap were deposited by electron beam deposition. Before the Cu deposition, an Ar-ion milling process was carefully conducted to the FM surface to obtain a clean interface between Cu and FM. The 200 nm wide and 40 nm thick Cu nanowires were deposited by Joule heat evaporator (also 30 nm, 50 nm Cu were employed in Cu thickness dependence experiments, see Supplementary Section 8). The 15 nm thick $Al_2O_3$ capping layers were deposited on the Cu nanowires by electron beam deposition. As clearly shown in the following sections, the local transport structure exhibits a great feasibility for measuring orbital accumulation distribution, offering a method for probing other phenomena of orbital accumulation. Our local measurement experiments certify the presence of orbital accumulation and provide a measurement on vertical orbital distribution.

**FM dependence**

To seek the corroboration of orbital accumulation, we conducted FM dependence experiments by employing various FMs, such as $Co_{25}Fe_{75}$, $Co_{50}Fe_{50}$, and $Ni_{81}Fe_{19}$. Several samples with the same geometry are measured as shown in Fig. S7c ~ Fig. S7h, suggesting the following relationship: $|2\Delta R^{(0)}|(Co_{25}Fe_{75}) > |2\Delta R^{(0)}|(Co_{50}Fe_{50}) \gg |2\Delta R^{(0)}|(Ni_{81}Fe_{19})$. The variations in $R_{DOEE}^{(0)}$ and $R_{IOEE}^{(0)}$ are caused by the magnetization switching in the electrodes, as indicated by the single-headed arrow pairs in the figures. The FM dependence experiments consistently correlate with the nonlocal measurements. The results show a strong dependence on FMs, suggesting that the measured signals originate from orbital response instead of spin response.

## Fig. S7 Local measurement of FM dependence

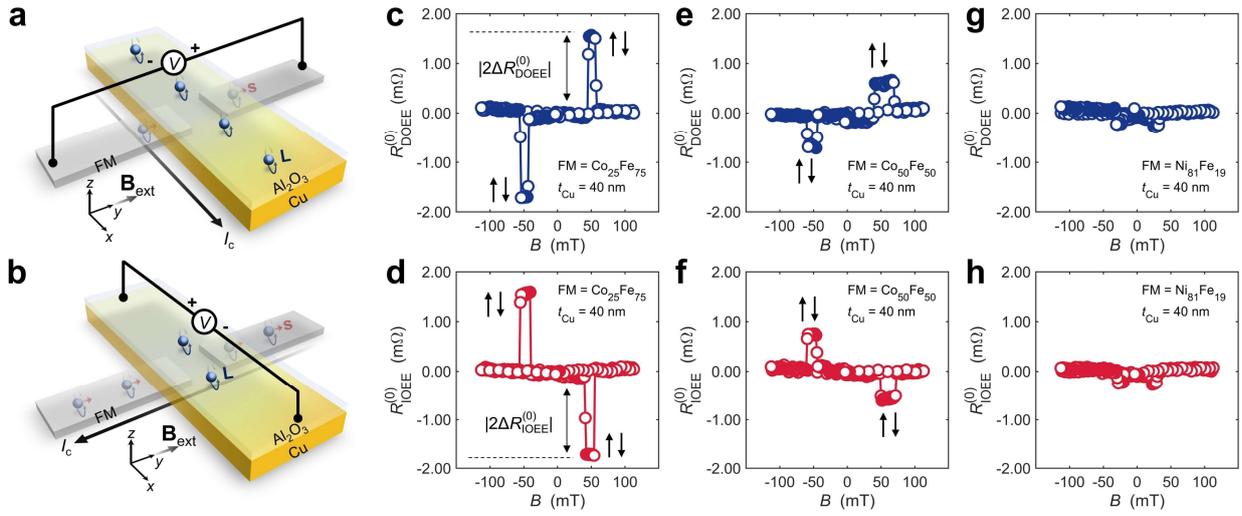

**Fig. S7 | Schematic illustrations of local transport structures and verification of orbital accumulation through ferromagnetic materials dependence experiment. a,** The local measurement configuration to observe DOEE (direct measurement). The nonequilibrium OAM are generated at $Al_2O_3$/$CuO_x$/Cu nanowire via DOEE. The orbital accumulation converts to spin accumulation in FM and induced a local orbital response. **b,** The local measurement configuration to observe IOEE (inverse measurement). The spin current is induced by charge accumulation, which further converts to orbital accumulation. The orbital accumulation then induces a charge current in $Al_2O_3$/$CuO_x$/Cu nanowire via IOEE. **c, d,** The local measurements signals of DOEE (**c**) and IOEE (**d**), with the Cu thickness $t_{Cu}$ = 40 nm and FM = $Co_{25}Fe_{75}$, showing that $|2\Delta R_{DOEE}^{(0)}| = |2\Delta R_{IOEE}^{(0)}| = 1.70$ m$\Omega$. The double-headed arrows represent the definition of $|2\Delta R_{DOEE}^{(0)}|$ and $|2\Delta R_{IOEE}^{(0)}|$. **e, f,** The local measurements result of DOEE (**e**) and IOEE (**f**), with the Cu thickness $t_{Cu}$ = 40 nm and FM = $Co_{50}Fe_{50}$, showing that $|2\Delta R_{DOEE}^{(0)}| = |2\Delta R_{IOEE}^{(0)}| = 0.80$ m$\Omega$. **g, h,** The local measurements result of DOEE (**g**) and IOEE (**h**), with the Cu thickness $t_{Cu}$ = 40 nm and FM = $Ni_{81}Fe_{19}$, whereas no OEE signals are shown. In **c** ~ **h**, the signals are globally offset to position their center at $R$ = 0 $\Omega$. The pairs of one-headed arrows represent the magnetization

configuration of FMs. The results show good agreement with Onsager's reciprocal relations.

**Section 5.    The role of spin current injected by electric current**

It is well established that charge-to-spin conversion is weak in systems composed of light elements due to the lack of strong spin-orbit coupling (SOC). Our device consists solely of $Al_2O_3$ and Cu ($CuO_x$) where the charge-to-spin conversion is expected to be negligible. We designed an additional experiment to provide clear validation. Here, we fabricated the new device by introducing an additional FM nanowire into the original device (see Fig. 1a in main text), as illustrated in Fig. S8a. This modification allows us to simultaneously measure the nonlocal OEE response and the nonlocal spin injection response within a single device.

The conventional nonlocal spin valve (NLSV) measurement configuration is shown in Fig. S8a. In this setup, an electric current $I_c$ is applied through the lower FM nanowire, while the voltage signal $V_{NLSV}$ is measured between the upper FM and the upper $Al_2O_3/CuO_x/Cu$ nanowire terminals. An external magnetic field is applied along the easy axis of FM. NLSV measurements were performed on two FM ($Ni_{81}Fe_{19}$ and $Co_{25}Fe_{75}$) at room temperature, with a separation distance of $\sim 400$ nm between the FMs and thickness of Cu 40nm. As shown in Fig. S8b and Fig. S8c, clear NLSV signals were observed in both $Ni_{81}Fe_{19}$ and $Co_{25}Fe_{75}$ devices, suggesting that spin current is injected into the $Al_2O_3/CuO_x/Cu$ nanowires.

The inverse orbital Edelstein effect measurement configuration is shown in Fig. S8d. In this setup, an electric current is applied through the lower FM nanowire, while the voltage signal $V_{IOEE}$ is measured between the two ends of $Cu_y$ terminals. The distance between $Cu_y$ and lower FM nanowire is $\sim 250$ nm. However, for IOEE measurements, those two FMs exhibited starkly contrasting responses. In devices with FM = $Ni_{81}Fe_{19}$, no characteristic IOEE response was detected, while in $Co_{25}Fe_{75}$ devices, a clear IOEE signal was observed. This pronounced FM dependence of the IOEE signal differs significantly from conventional spin responses, which are typically less sensitive to the choice of FM. Based on these results, we attribute the observed signal, as shown in the main text, predominantly to an orbital response, while the contribution of spin injection is minimal.

**Fig. S8**

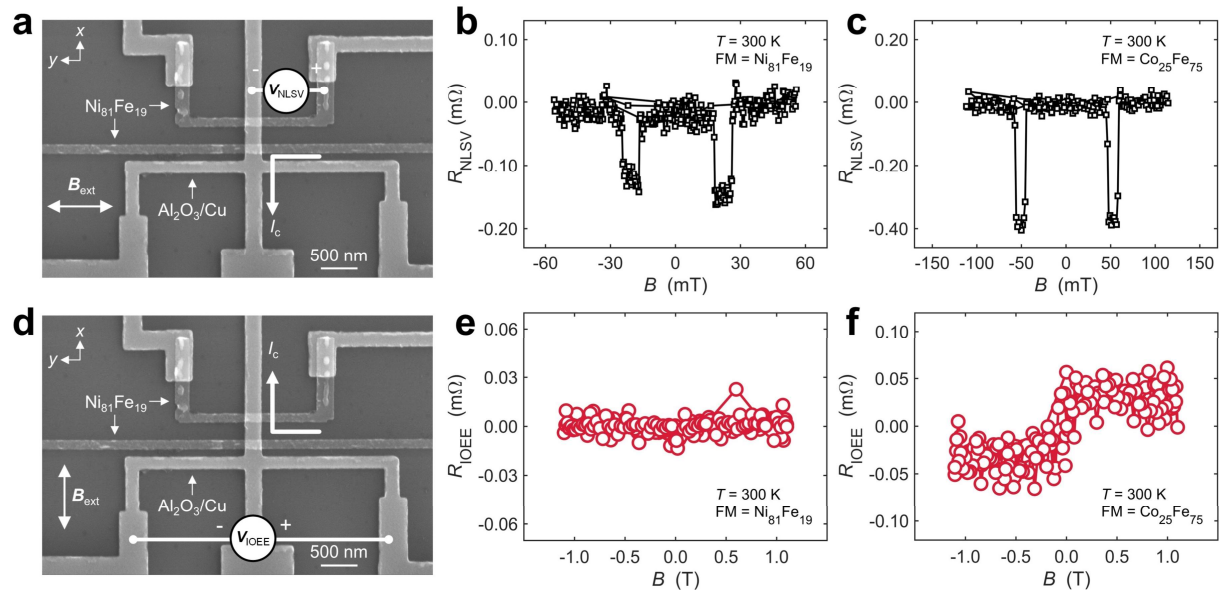

**Fig. S8 | Nonlocal spin valve measurement and inverse OEE measurement at room temperature. a,** The NLSV measurement configuration. The spin current is injected into the $Al_2O_3/CuO_x/Cu$ nanowire. **b,** The NLSV measurement results of device with $Ni_{81}Fe_{19}$ at room temperature. **c,** The NLSV measurement results of device with $Co_{25}Fe_{75}$ at room temperature. **d,** The nonlocal IOEE measurement configuration. **e,** The nonlocal IOEE measurement of device with $Ni_{81}Fe_{19}$ at room temperature. No characteristic signal is observed. **f,** The nonlocal IOEE measurement of device with $Co_{25}Fe_{75}$ at room temperature. A typical nonlocal IOEE signal is observed.

**Section 6.    The nonlocal orbital distribution model which considers bypass current**

**The COMSOL simulation of bypass current density**

We first performed the COMSOL simulation to study the flow of charge current density, as shown in Fig. S9a. The $y$-component of bypass current density $j_{\mathrm{by}}^{y}$ is evaluated (averaged value) over the cross section of $Cu_x$ nanowire above the center of FM nanowires at various distance. $j_{\mathrm{by}}^{y}$ as a function of $d$ is summarized in Fig. S9b. The selection of $d$ was based on practical devices. Assuming the $j_{\mathrm{by}}^{y}$ decays exponentially as $j_{\mathrm{by}}^{y} \propto \exp\left(-\frac{d}{\lambda_{\mathrm{by}}}\right)$, the fitting yields a decay length of bypass current $\lambda_{\mathrm{by}} \cong 47$ nm (Fig. S2c). However, fitting the results of experimental nonlocal transport measurement to $\Delta R_{\mathrm{NL}} \propto \exp\left(-\frac{d}{\lambda_{\mathrm{NL}}}\right)$ gives rise to a lateral decay length $\lambda_{\mathrm{NL}} \cong 110$ nm, as shown in the Fig. S9c and Fig. S9d. The stimulated results of bypass current show a decay two times faster than observed signals and suggest that the bypass current is not the main origin of the signals. Therefore, we believe the signals are the combination of bypass effect of charge current and the lateral distribution of OAM.

**The analytical model of nonlocal distribution of orbital accumulation considering bypass current**

We hereby analyze $\lambda_{\mathrm{o}}$ by considering the combined contribution of the bypass effect of charge current and the nonlocal distribution of orbital accumulation. We assume the $y$-component charge current density is uniform in the Cu cross region, which can be expressed as:

$$j_{\mathrm{c}}^{y}(0) = \frac{I_{\mathrm{c}}\xi_{\mathrm{by}}}{W_y t_{\mathrm{Cu}}}. \qquad (S6-1)$$

where $I_{\mathrm{c}}$ is the applied current, $\xi_{\mathrm{by}}$ is a shutting constant, $W_y$ is the width of $Cu_y$ nanowire, $t_{\mathrm{Cu}}$ is the thickness of the Cu. To provide a simple picture of the charge bypass and orbital decay joint effect, we consider an effective one-dimensional model, where the $y$-component of bypass current density $j_{by}^y$ is solely a function of the distance $x$ from the center of Cu$_y$ nanowire (as shown in the Fig. S3a). Its magnitude can be written as[4,5]:

$$j_{by}^y(x) = j_c^y(0) \exp\left(-\frac{\pi x}{W_x}\right), \qquad (S6-2)$$

where $W_x$ is the width of the Cu$_x$ nanowire. $\xi_{by}$ can be self-consistently determined as follows:

$$2t_{Cu} \int_0^{+\infty} j_{by}^y(x)dx = 2t_{Cu} \int_0^{+\infty} \frac{I_c \xi_{by}}{W_y t_{Cu}} \exp\left(-\frac{\pi x}{W_x}\right)dx = I_c,$$

$$\xi_{by} = \left(\frac{\pi W_y}{2W_x}\right). \qquad (S6-3)$$

Here, the influence of FM (because the resistance of FM nanowire is much more significant than that of Cu nanowire) and the difference in width between Cu$_y$ and Cu$_x$ are ignored. Factor 2 is included since the bypass effect has two sides. The generated orbital accumulation $a_o^x$ can be induced by $j_{by}^y$ at $x$ as following ($y$-polarized OAM cancel out each other, as shown in the Fig. S3b):

$$a_o^g(x) = q_{ICO} t_{CuO} j_{by}^y(x) \qquad (S6-4)$$

where $t_{CuO}$ is the interfacial oxidization thickness, and $q_{ICO}$ is the interfacial charge-to-orbital conversion efficiency similar to the spin Rashba counterpart in topological insulator[8–10]. Subsequently, $a_o^g(x)$ undergoes a decay from $x$ to $d$ (from its generation location to the center of FM). Thus, the contribution from $a_o^g(x)$ to the orbital accumulation at $d$ can be expressed as following if we assume it decays exponentially (as shown in the Fig. S3c):

$$\Delta\alpha_o^d(d,x) = a_o^g(x) \exp\left(-\frac{d-x}{\lambda_o}\right) \qquad (S6-5)$$

where $\lambda_o$ is the lateral decay length of orbital accumulation. The overall orbital accumulation $\alpha_o^x$ at $d$ is an integration of $\Delta\alpha_o^d(d,x)$ over the area from the center of Cu cross to the center of FM given by:

$$\alpha_o^x(d) = \int_0^d \int_{-W_x/2}^{+W_x/2} \Delta\alpha_o^d(d,x)\, dx\, dy$$

$$= \int_0^d \int_{-W_x/2}^{+W_x/2} q_{ICO} t_{CuO} \frac{I_c}{t_{Cu}} \frac{\pi}{2W_x} \exp\left(-\frac{\pi x}{W_x}\right) \exp\left(-\frac{d-x}{\lambda_o}\right) dx\, dy$$

$$= \frac{\pi I_c q_{ICO} t_{CuO}}{2 t_{Cu}\left(\frac{1}{\lambda_o} - \frac{\pi}{W_x}\right)}\left[\exp\left(-\frac{\pi d}{W_x}\right) - \exp\left(-\frac{d}{\lambda_o}\right)\right]. \qquad (S6-6)$$

The nonlocal voltage generated from the chemical potential is proportional to $\alpha_o^x(d)$. Suppose $V = A_o \alpha_o^x(d)$, where the coefficient $A_o$ depends on the thickness of the Cu layer, the orbital polarization of the FM, the interfacial orbital transmission efficiency and the area of the FM/Cu junction, the nonlocal direct orbital Edelstein resistance can then be described as:

$$2\Delta R_{DOEE} = \frac{V}{I_c} = \frac{\pi q_{ICO} t_{CuO}}{2 t_{Cu}\left(\frac{1}{\lambda_o} - \frac{\pi}{W_x}\right)}\left[\exp\left(-\frac{\pi d}{W_x}\right) - \exp\left(-\frac{d}{\lambda_o}\right)\right]$$

$$= \frac{A}{\left(\frac{1}{\lambda_o} - \frac{\pi}{W_x}\right)}\left[\exp\left(-\frac{\pi d}{W_x}\right) - \exp\left(-\frac{d}{\lambda_o}\right)\right], \qquad (S6-7)$$

where $2\Delta R_{DOEE}$ represents the overall change of nonlocal direct orbital Edelstein effect, and $A$ is the fitting parameter containing the orbital Edelstein length $q_{ICO}$ and the orbital absorption efficiency $A_o$.

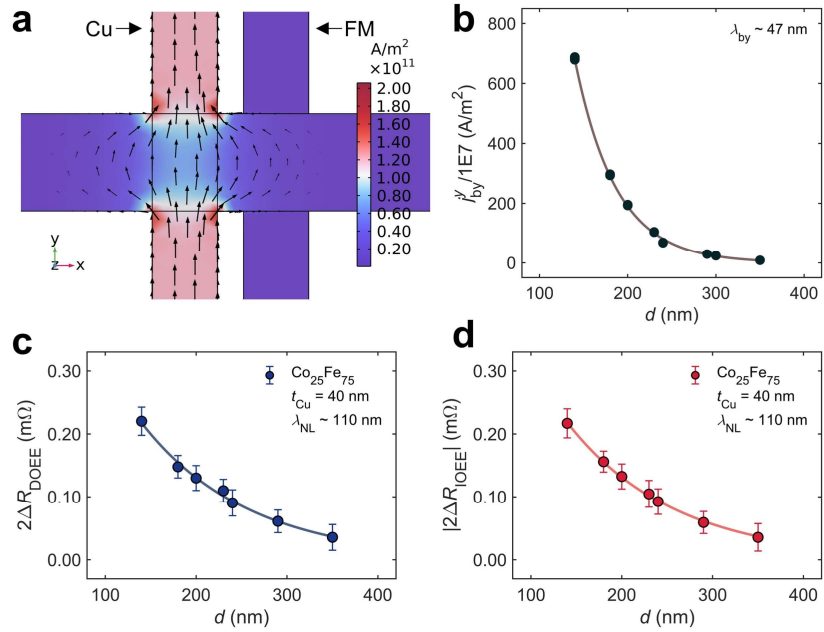

**Fig. S9 | The analysis of bypass current. a,** The COMSOL simulation for $j_{by}^{y}$ in our nonlocal transport structure, where FM nanowire (right) is under Cu nanowire (left). The color bar denotes the magnitude of the $y$-component current density ($j_{by}^{y}$). The black arrows denote the flow of the charge current density ($j_c$) while their lengths indicate the relative magnitude. **b,** The exponential fitting results of $j_{by}^{y}$ (the average over the cross-sectional area) to $d$, suggesting a decay length of bypass current of about 47 nm. **c, d,** The exponential fitting ($2\Delta R = A\exp(-d/\lambda_o)$) result of $2\Delta R_{DOEE}$ (**c**) and $|2\Delta R_{DOEE}|$ (**d**) measured by nonlocal transport structure, indicating a characteristic length of nonlocal resistance about 110 nm. The distinction implies that bypass current is not the main origin of our nonlocal signal. The error bars indicate the noise level in raw data.

**Section 7.    The analysis of the angle dependence**

Here, we discuss the deviation between experimental data and the cosine curves shown in Fig. 2c and 2d in the main text. The deviation can be explained if we make the following two assumptions:

1.  Angular dependence is governed by the direction of magnetization **m** rather than the external magnetic field $\mathbf{B}_{\text{ext}}$. Thus, $2\Delta R_{\text{DOEE}}$ and $2\Delta R_{\text{IOEE}}$ can be written as:

$$2\Delta R_{\text{DOEE}} = |2\Delta R_{\text{IOEE}}| = f_1 \cos(\phi_m) \qquad (S7-1)$$

    where the $\phi_m$ denotes the in-plane angle of **m** to the *x*-axis.

2.  Ferromagnet (FM) has an easy axis along the *y*-axis. Therefore, we have:

$$F = -\mathbf{B}_{\text{ext}} \cdot \mathbf{m} - K(\mathbf{m} \cdot \hat{y})^2, \qquad (S7-2)$$

    where $F$ denotes the free energy density of FM and $K$ denotes the in-plane easy-axis anisotropy along the *y*-axis.

Thus, Assumption 1 implies that $2\Delta R_{\text{OEE}}$ depends on $\phi_m$ rather than $\Phi$ (the angle of $\mathbf{B}_{\text{ext}}$ from the *y*-axis). When $\mathbf{B}_{\text{ext}}$ is sufficiently strong, $\phi_m$ will be close to $\Phi$, but there will be a slight deviation between $\phi_m$ and $\Phi$, which decays with increasing $\mathbf{B}_{\text{ext}}$. The in-plane magnetic anisotropy will determine the amount of deviation. In Assumption 2, the difference between $\phi_m$ and $\Phi$ can be determined by minimizing $F$ with respect to $\phi_m$. For minimization, we rewrite Eq. S7−2 as follows:

$$F = -\mathbf{B}_{\text{ext}} \cdot \mathbf{m} - K(\mathbf{m} \cdot \hat{y})^2 = -B_{\text{ext}} \cos(\phi_m - \Phi) - K \sin^2 \phi_m. \qquad (S7-3)$$

From $\partial F / \partial \phi_m = 0$, we obtain:

$$B_{\text{ext}} \sin(\phi_m - \Phi) = K \sin 2\phi_m. \qquad (S7-4)$$

When $B_{\text{ext}}$ is much larger than $K$, we may take:

$$\phi_m = \Phi + \delta\phi_m. \qquad (S7-5)$$

By inserting Eq. S7−5 into Eq. S7−3, one obtains:

$$\delta\phi_m = \frac{K}{B_{\text{ext}}}\sin 2\Phi + O\left(\frac{K}{B_{\text{ext}}}\right)^2. \qquad (S7-6)$$

From Eq. S7-1, S7-5 and S7-6, one obtains:

$$2\Delta R_{\text{DOEE}} = |2\Delta R_{\text{IOEE}}| = f_1 \cos\left(\Phi + \frac{K}{B_{\text{ext}}}\sin 2\Phi\right) \qquad (S7-7)$$

which can be modified as

$$\begin{aligned}
2\Delta R_{\text{DOEE}} = |2\Delta R_{\text{IOEE}}| &= f_1 \cos\left(\Phi + \frac{|f_2|}{2f_1}\sin 2\Phi\right) \\
&= f_1 \cos\left(\Phi - \frac{f_2}{f_1}\sin\Phi\cos\Phi\right) \\
&\approx f_1 \left(\cos\Phi + \frac{f_2}{f_1}\sin^2\Phi\cos\Phi\right) \\
&= f_1\cos\Phi + f_2\sin^2\Phi\cos\Phi \qquad (S7-8)
\end{aligned}$$

if the ratio $|f_2|/f_1$ satisfies the following relation,

$$\frac{|f_2|}{2f_1} = \frac{K}{B_{\text{ext}}}. \qquad (S7-9)$$

Fitting the deviation value $2\Delta R_{\text{OEE}} - 2\Delta R_{\text{fit}}$ to Eq. S7-8 shows a high consistency, as shown in Fig. S10a and Fig. S10b, where $K$ is estimated as $0.20 \sim 0.26$ T with Eq. S7-9. The value is reasonable for 100 nm wide and 20 nm thick $Co_{25}Fe_{75}$ nanowires, implying that the deviation may originate from the magnetization anisotropy of FM.

In addition to magnetization anisotropy, orbital anisotropy may also contribute to the deviation via the orbital-spin conversion process. Eq. S7-8 can be rewritten as:

$$2\Delta R_{\text{DOEE}} = |2\Delta R_{\text{IOEE}}| = (f_1 + f_2)\mathbf{m}\cdot\mathbf{B} - f_2(\mathbf{m}\cdot\mathbf{B})^2\mathbf{m}\cdot\mathbf{B}, \qquad (S7-10)$$

where the $f_2$ term is a high-order contribution in the orbital-spin conversion that comes from the orbital anisotropy. Similar orbital anisotropy was reported in an orbital torque study, where the torque efficiency shows a sizable high-order term in the angular dependence of the orbital torque generated by OEE[11]. As both mechanisms may contribute to the $f_2$ term, the observed slight deviation in the angular dependence is reasonable.

**Fig. S10**

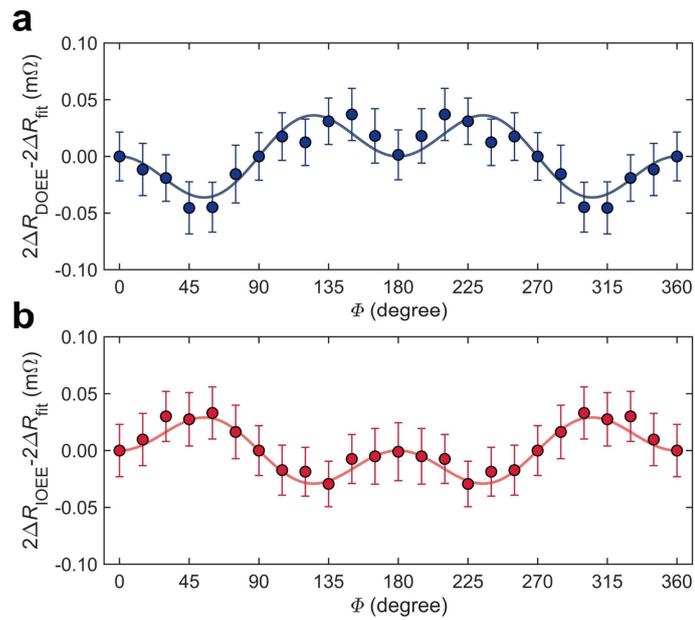

**Fig. S10 | The deviation of angle dependence data. a, b,** The fitting of the direct measurement (**a**) and inverse measurement (**b**) data to Eq. S7 -8, suggesting the magnetization anisotropy may lead to the deviation in Fig. 2c and 2d (in main text).

**Section 8.    Cu thickness dependence measured in local transport structure**

We conducted experiments to examine the dependency of Cu thickness by using local transport measurement. The samples measured share the same geometry except the Cu thickness. $R_{\text{DOEE}}^{(0)}$ and $R_{\text{IOEE}}^{(0)}$ for Cu thicknesses of 30 nm, 40 nm, and 50 nm are shown in Fig. S11a ~ Fig. S11f, while $2\Delta R_{\text{DOEE}}^{(0)}$ and $2\Delta R_{\text{IOEE}}^{(0)}$ are summarized in Fig. S11g and Fig. S11h. The signals $2\Delta R_{\text{DOEE}}^{(0)}$ and $2\Delta R_{\text{IOEE}}^{(0)}$ decrease as Cu thickness increases because of the longer transport distance, which is consistent with the results obtained in nonlocal measurements. By assuming the orbital accumulation distribution decays exponentially in local transport, the signal can be expressed as $|2\Delta R_{\text{DOEE}}^{(0)}| = |2\Delta R_{\text{IOEE}}^{(0)}| = A_{\text{Cu}}^{(0)} \exp(-t_{\text{cu}}/\lambda_{\text{o}}^{z})$. Fitting the data to the equation yields that $\lambda_{\text{o}}^{z} \sim 36\,\text{nm}$, which is consistent with the value ($\sim 25\,\text{nm}$) obtained from the nonlocal Cu thickness dependence experiment.

**Fig. S11    Local measurement of Cu thickness dependence**

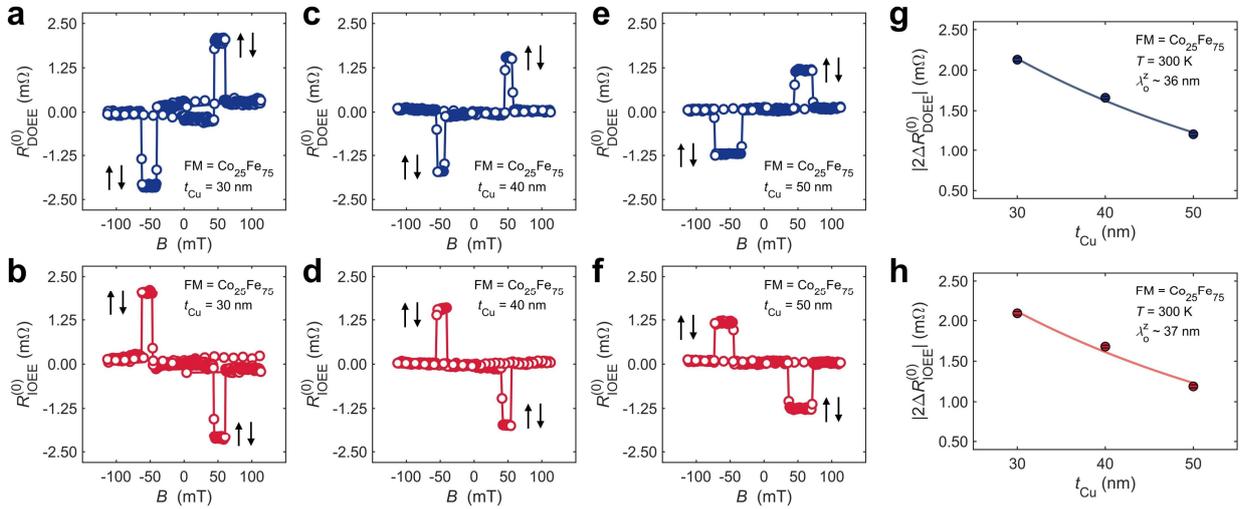

**Fig. S11 | Cu thickness dependence experiment exploiting local transport structures.**
**a, b** The local transport measurement results of $R_{\text{DOEE}}^{(0)}$ and $R_{\text{IOEE}}^{(0)}$ with Cu thickness of 30 nm, showing that $|2\Delta R_{\text{DOEE}}^{(0)}| = |2\Delta R_{\text{IOEE}}^{(0)}| = 2.20\,\text{m}\Omega$. **c, d** The results of $R_{\text{DOEE}}^{(0)}$ and $R_{\text{IOEE}}^{(0)}$ with Cu thickness of 40 nm, showing that $|2\Delta R_{\text{DOEE}}^{(0)}| = |2\Delta R_{\text{IOEE}}^{(0)}| = 1.70\,\text{m}\Omega$. **e, f** The results of $R_{\text{DOEE}}^{(0)}$ and $R_{\text{IOEE}}^{(0)}$ with Cu thickness of 50 nm, showing that $|2\Delta R_{\text{DOEE}}^{(0)}| = |2\Delta R_{\text{IOEE}}^{(0)}| = 1.20$ mΩ. The results are consistent with nonlocal measurement results and show good agreement with Onsager's reciprocal relations. In **a ~ f**, the signals are globally offset to position their center at $R = 0$ Ω. The pairs of one-headed arrows represent the magnetization configuration of FMs. **g, h,** $|2\Delta R_{\text{DOEE}}^{(0)}|$ and $|2\Delta R_{\text{IOEE}}^{(0)}|$ with various Cu thickness. The error bars indicating the noise level of raw data are smaller than the range of markers. The solid curves demonstrate the fitting of data to the exponential decay equation, showing a local decay length of orbital accumulation ~ 36 nm. The measurements are performed under room temperature. The results show good agreement with Onsager's reciprocal relations.

**Section 9.      Temperature dependence measured in local transport structure**

Experimental results for nonlocal DOEE and IOEE measurements on sample A ($d = 140$ nm, $t_{Cu} = 40$ nm, FM = $Co_{25}Fe_{75}$) as a function of temperature are presented in Fig. S12a and Fig. S12b, respectively.

**Fig. S12**

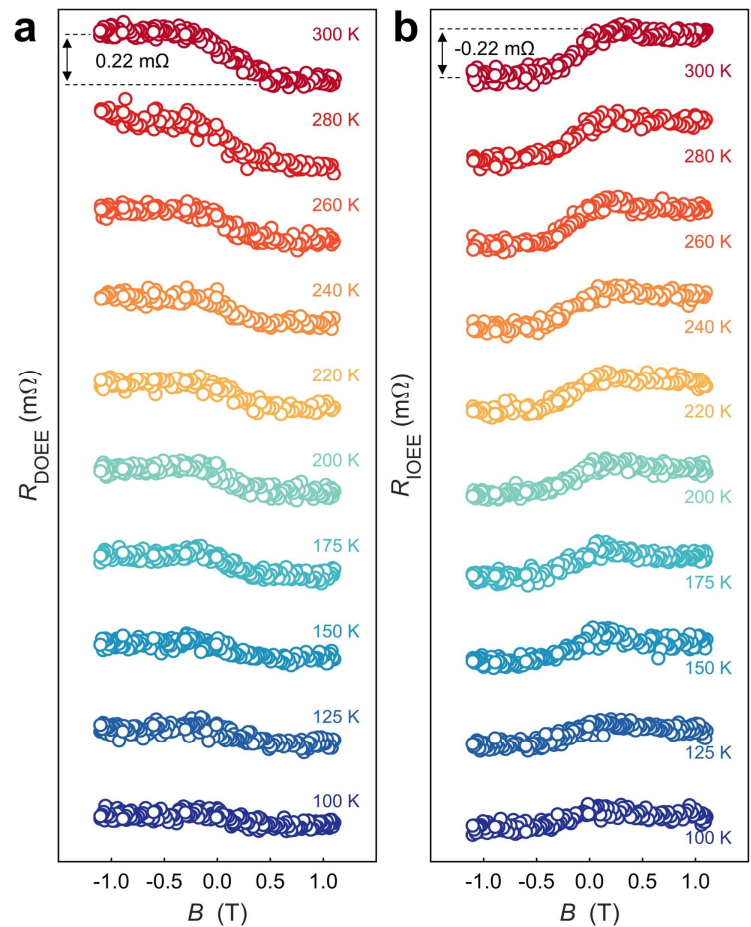

**Fig. S12 | Temperature dependence measurement exploiting nonlocal transport structures** (FM = $Co_{25}Fe_{75}$). **a, b** The results of $R_{DOEE}$ (**a**) and $R_{IOEE}$ (**b**) at various temperatures measured in sample A, respectively. As temperature decreases, the orbital response diminishes. The temperature dependence of nonlocal orbital transport is consistent with local measurement results and shows good agreement with Onsager's reciprocal relations.

We explored the temperature dependence of samples with 40 nm thick Cu employing $Co_{25}Fe_{75}$ through local OEE measurement. The results at different temperatures are shown in Fig. S13a and Fig. S13b. A consistent reduction in $R_{\mathrm{DOEE}}^{(0)}$ and $R_{\mathrm{IOEE}}^{(0)}$ with decreasing temperature was observed, aligning with results from nonlocal measurement experiments, showing a special temperature dependence of orbital accumulation distribution.

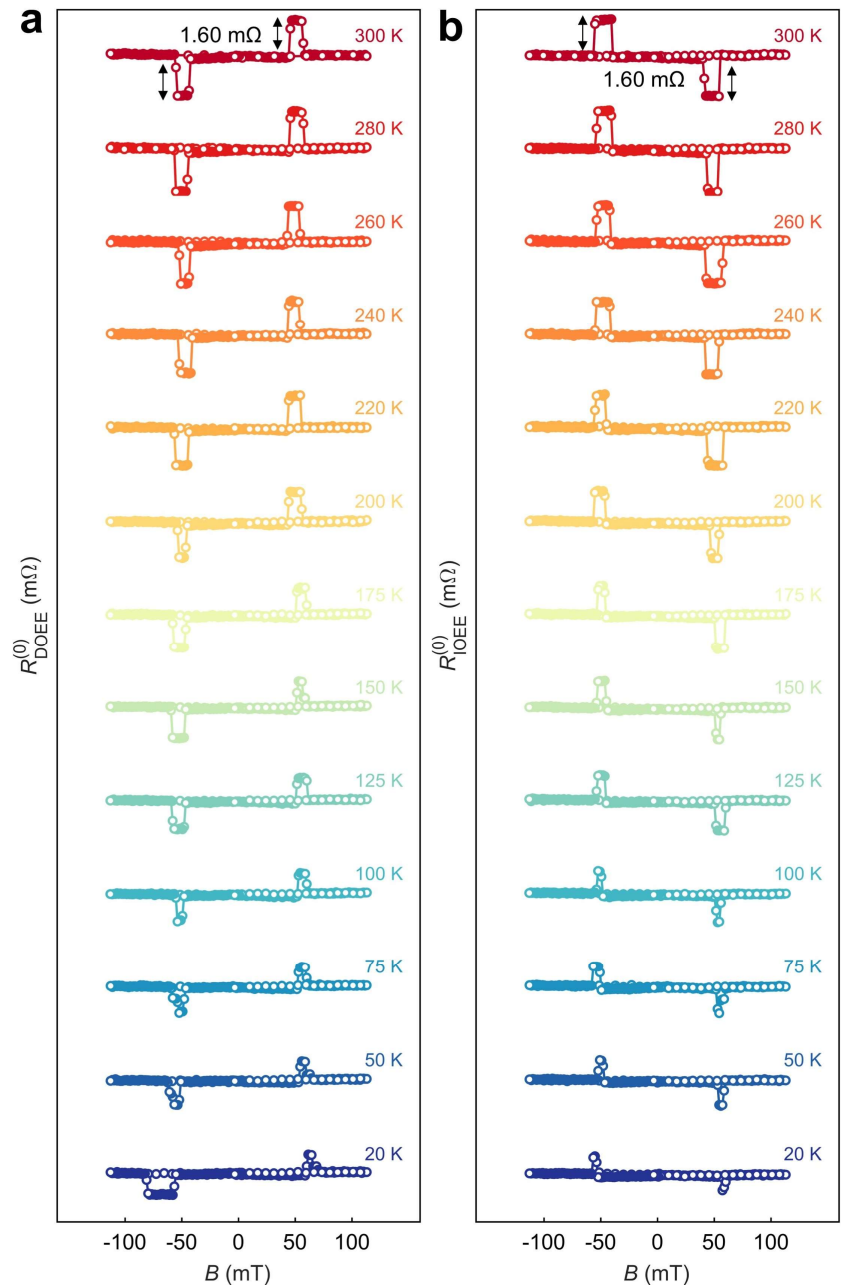

**Fig. S13 | Temperature dependence experiment exploiting local transport structures (FM = Co₂₅Fe₇₅). a, b** The results of $R_{DOEE}$ (**a**) and $R_{IOEE}$ (**b**) at various temperatures for Cu thickness of 40 nm, respectively. The temperature dependence of local orbital transport is consistent with nonlocal measurement results and shows good agreement with Onsager's reciprocal relations.

Here, the IOEE temperature dependence measurement results ($2\Delta R_{\mathrm{IOEE}}$ and $\left|2\Delta R_{\mathrm{IOEE}}^{(0)}\right|$ in

Fig. S12b and Fig. S13b, respectively) are further summarized in the Fig. S14.

**Fig. S14**

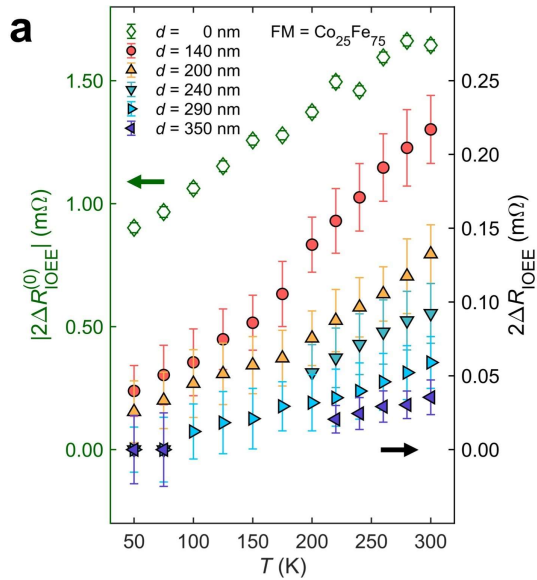

**Fig. S14 | Temperature dependence of IOEE** (FM = Co$_{25}$Fe$_{75}$). **a,** The results of

$\left|2\Delta R_{\mathrm{IOEE}}^{(0)}\right|$ and $2\Delta R_{\mathrm{IOEE}}$ as a function of temperature. The green open diamonds indicate the $\left|2\Delta R_{\mathrm{IOEE}}^{(0)}\right|$. The circles and triangles markers filled with other color indicate the

$\left|2\Delta R_{\mathrm{IOEE}}\right|$. The error bars indicate the noise level in the raw data.

Furthermore, we studied the temperature dependence of DOEE (Fig. S15a) and IOEE (Fig. S15b) in the $Ni_{81}Fe_{19}$ devices. In all temperature points, no characteristic signal is shown.

**Fig. S15**

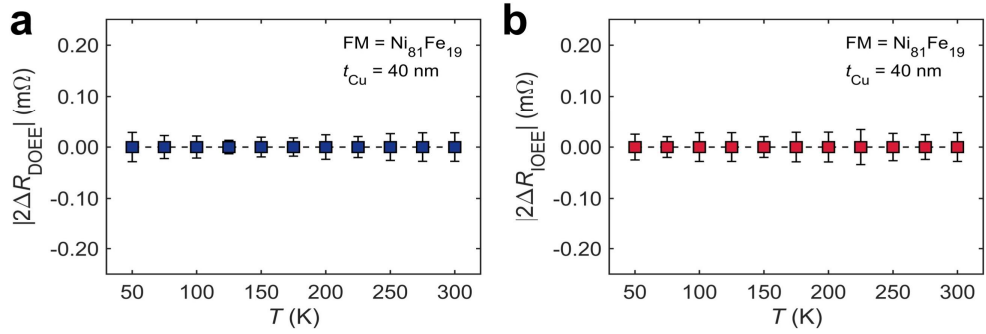

**Fig. S15 | Temperature dependence of IOEE measured in FM = $Ni_{81}Fe_{19}$ device. a,** $|2\Delta R_{DOEE}|$ as a function of temperature. **b,** $|2\Delta R_{IOEE}|$ as a function of temperature. The dashed line represents the position $2\Delta R = 0$. The error bars indicate the noise level in the raw data.

The IOEE data obtained from different devices at the same temperature were fitted using Eq. 1 to obtain the temperature dependent $\lambda_o$ ($\lambda_o$ is the only fitting parameter), with the fitting process shown in Fig. S16.

**Fig. S16**

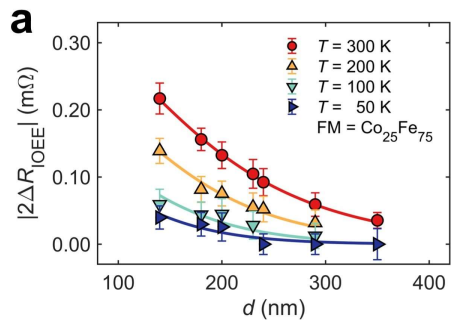

**Fig. S16 | The fitting of nonlocal IOEE data. a,** The fitting of the $2\Delta R_{IOEE}(T)$ data with Eq. (1). The error bars indicate the noise levels of $R_{DOEE}$.

**Section 10.    The resistivity of Cu and Co$_{25}$Fe$_{75}$ and the temperature dependence of bypass current**

The resistivities of Cu and Co$_{25}$Fe$_{75}$ at various temperatures were measured using four-terminal methods. The Cu is 100 nm wide and 40 nm thick. The Co$_{25}$Fe$_{75}$ is 100 nm wide and 20 nm thick. The results are shown in Fig. S17.

**Fig. S17**

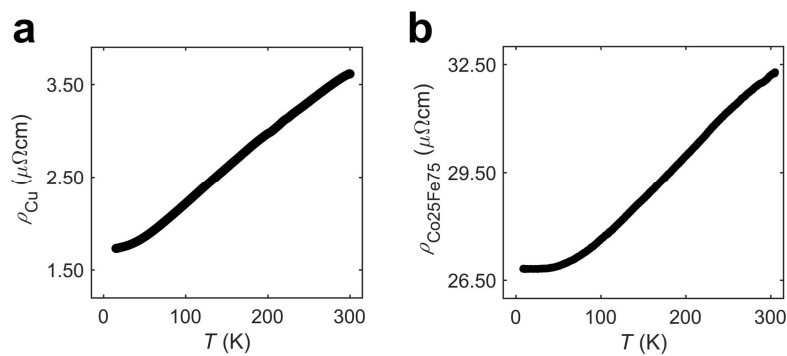

**Fig. S17 | Temperature dependence of resistivity of Cu and Co$_{25}$Fe$_{75}$. a, b,** The resistivity of Cu (**a**) and Co$_{25}$Fe$_{75}$ (**b**) at various temperature measured using four probe measurements.

As discussed in Supplementary Section 6, the diffusion length of the bypass current in our device is approximately 47 nm at room temperature, while the decay length of orbital accumulation exceeds 100 nm. To conclusively exclude the influence of bypass current on our estimation of the orbital accumulation decay length, we performed low-temperature COMSOL simulations to assess the temperature dependence of the bypass current. Similarly, we simulated the surface averaged value of $j_{by}^{y}$ across the cross section of the Cu$_x$ nanowire above the FM nanowires at various distances. The experimentally obtained temperature-dependent conductivities of Cu (Fig. S17a) and FM (Fig. S17b) were used as inputs to simulate the bypass current at various temperatures. The simulation result at 50 K is shown in Fig. S18a suggesting an unchanged value of $j_{by}^{y}$ at various temperature, and the diffusion length of bypass current $\lambda_{by}$ as a function of temperature are shown in Fig. S18b. $\lambda_{by}$ shows no temperature dependence, which is consistent with previous research[4,5], while $\lambda_{o}$ decreases at lower temperatures, suggesting that the observed temperature dependent decay behavior is primarily attributed to the orbital response and is not significantly influenced by the presence of bypass current. This is because the resistivity of FM is always one order larger than that of Cu, and the bypass current can be regarded as only flowing in the Cu layer. In a single material, as the resistivity is homogenous, the bypass current distribution is determined merely by the geometric of the device[4,5].

**Fig. S18**

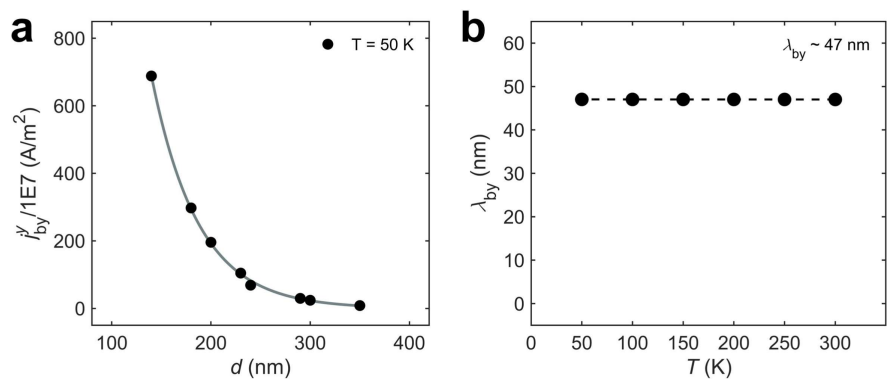

**Fig. S18 | Temperature dependence of bypass current. a,** The bypass current as a function of distance under 50 K. Fitting the data to exponent decay equation (grey curve) gives a diffusion length of bypass current $\lambda_{by}$ about 47 nm, which is the same as the $\lambda_{by}$ at 300 K. **b,** The temperature dependence of the diffusion length bypass current. $\lambda_{by}$ remains 47 nm at various temperatures.

Furthermore, based on the measured resistivity we calculated the mean free path of electrons in our devices. The mean free path of electron ($l_e$) in Cu can be calculated as follows:

$$l_e = v_F \cdot \tau, \qquad\qquad (S10-1)$$

where $v_F$ is the electron velocity at the Fermi surface, $\tau$ is the relaxation time.

According to Drude model and Ohm's law, $\tau$ can be calculated as follows:

$$\tau = \frac{m_e}{\rho n e^2}, \qquad\qquad (S10-2)$$

where $m_e$ is the mass of electron, $\rho$ is the resistivity of Cu, $n$ is the free electron density of Cu, and $e$ is the electron charge. Taking the parameters[12] that $n = 8.49 \times 10^{28} \ \mathrm{m^{-3}}$ and $v_F = 1.57 \times 10^8 \mathrm{cm/s}$, we obtained that $l_e(T = 300\mathrm{K}) = 19 \ \mathrm{nm}$ and $l_e(T = 50\mathrm{K}) = 38 \ \mathrm{nm}$.

**Section 11.     Temperature-dependent multiple-step hopping**

We consider the hopping between the states in the oxidized Cu, which can be mediated by the metallic Cu, as illustrated in Fig. S19a. We assume that for each hopping between two neighboring grains, the probability for an electron to maintain its OAM information is

$$P = C \exp(-\Delta E/k_B T) = \exp(-\alpha - \Delta E/k_B T), \qquad (S11-1)$$

where $\alpha = -\ln C$ is a parameter related to the wavefunction overlapping, $k_B$ is the Boltzmann constant, and $\Delta E$ is the level mismatch between the two neighboring grains. This $\Delta E$ is related to the inelastic processes that suppress the OAM. The OAM-maintaining hopping probability across $N$ grains is

$$P(N) = \exp(-N\alpha - N\Delta E/k_B T), \qquad (S11-2)$$

and the total hopping distance is $d = Nr$, where $r$ is the grain size. Thus, we can rewrite the hopping probability as

$$P(N) = \exp[-\alpha d/r - d\Delta E/(rk_B T)] = \exp(-d/\lambda_H), \qquad (S11-3)$$

where

$$\lambda_H = \frac{r}{\alpha + \Delta E/k_B T}. \qquad (S11-4)$$

Taking $\Delta E/k_B = 10$ K, $r = 10$ nm, $\alpha = 1/15$, one can obtain a temperature dependence of $\lambda_H$ (Fig. S19b), which is comparable with the experimental $\lambda_o$ as shown in Fig. 5b in main text.

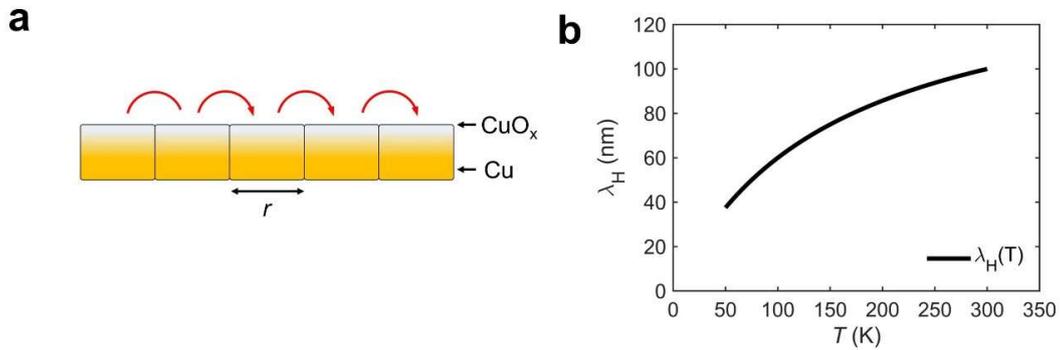

**Fig. S19 | a,** Schematic of the multiple hopping in Cu/CuO$_x$ bilayer. **b,** Temperature dependence of $\lambda_H$.

Such a hopping language description corresponds to that when conductive bands involving both Cu and oxidized Cu are formed continuously across distance $\lambda_H$, the eigenstates with OAM can be created and the OAM propagation is allowed across a similar distance.